\documentclass{article}

\usepackage[utf8]{inputenc}
\usepackage{authblk}
\usepackage{url}
\usepackage{amsthm,amsmath}
\RequirePackage[numbers]{natbib}
\RequirePackage{hyperref}
\usepackage[utf8]{inputenc} %unicode support
\usepackage{graphicx,subfigure}
\usepackage{float} 
\usepackage{url}
\usepackage{multicol}
\usepackage{caption}
\usepackage{subfigure}
\makeatletter
\renewcommand{\fnum@figure}{Fig.\thefigure}
\makeatother
\usepackage[hypcap=false]{caption}
\usepackage{hyperref}

\title{Improving Classification Model Performance on Chest X-Rays through Lung Segmentation}

\author[1]{Hilda Azimi\thanks{Hilda.Azimi@nrc-cnrc.gc.ca}}   
\author[2]{Jianxing Zhang\thanks{Jianxing.Zhang@uwaterloo.ca}} 
\author[1]{Pengcheng Xi\thanks{Pengcheng.Xi@nrc-cnrc.gc.ca}} 
\author[3]{Hala As'ad\thanks{Asad.Hala@gmail.com}} 
\author[1]{Ashkan Ebadi\thanks{Ashkan.Ebadi@nrc-cnrc.gc.ca}} 
\author[1]{St\'ephane Tremblay\thanks{St\'ephane.Tremblay@nrc-cnrc.gc.ca}} 
\author[2]{Alexander Wong\thanks{Alexander.Wong@uwaterloo.ca}} 

\affil[1]{National Research Council Canada}
\affil[2]{University of Waterloo}
\affil[3]{Ford Motor Company}
\date{\vspace{-5ex}}

\begin{document}

\maketitle

\begin{abstract}
Background: %if any
Chest radiography is an effective screening tool for diagnosing pulmonary diseases. In computer-aided diagnosis, extracting the relevant region of interest, i.e., isolating the lung region of each radiography image, can be an essential step towards improved performance in diagnosing pulmonary disorders. 

Methods: %if any
In this work, we propose a deep learning approach to enhance abnormal chest x-ray (CXR) identification performance through segmentations. Our approach is designed in a cascaded manner and incorporates two modules: a deep neural network with criss-cross attention modules (XLSor) for localizing lung region in CXR images and a CXR classification model with a backbone of a self-supervised momentum contrast (MoCo) model pre-trained on large-scale CXR data sets. The proposed pipeline is evaluated on ``Shenzhen Hospital (SH) data set" for the segmentation module, and ``COVIDx data set" for both segmentation and classification modules. Novel statistical analysis is conducted in addition to regular evaluation metrics for the segmentation module. Furthermore, the results of the optimized approach are analyzed with gradient-weighted class activation mapping (Grad-CAM) to investigate the rationale behind the classification decisions and to interpret its choices.

Results and Conclusion: %if any
Different data sets, methods, and scenarios for each module of the proposed pipeline are examined for designing an optimized approach, which has achieved an accuracy of 0.946 in distinguishing abnormal CXR images (i.e., Pneumonia and COVID-19) from normal ones. Numerical and visual validations suggest that applying automated segmentation as a pre-processing step for classification improves the generalization capability and the performance of the classification models. 
\end{abstract}


%%%%%%%%%%%%%%%%%%%%%%%%%%%%%%%%%%%%%%%%%%%%%%%%
%%                                            %%
%% The Main Body begins here                  %%
%%                                            %%
%% Please refer to the instructions for       %%
%% authors on:                                %%
%% https://www.biomedcentral.com/getpublished %%
%% and include the section headings           %%
%% accordingly for your article type.         %%
%%                                            %%
%% See the Results and Discussion section     %%
%% for details on how to create sub-sections  %%
%%                                            %%
%% use \cite{...} to cite references          %%
%%  \cite{koon} and                           %%
%%  \cite{oreg,khar,zvai,xjon,schn,pond}      %%
%%                                            %%
%%%%%%%%%%%%%%%%%%%%%%%%%%%%%%%%%%%%%%%%%%%%%%%%

%%%%%%%%%%%%%%%%%%%%%%%%% start of article main body
% <put your article body there>

%%%%%%%%%%%%%%%%
%% Background %%
%\begin{multicols}{2}
\section*{Introduction}
Computer-aided analysis of chest x-ray (CXR) images assists radiologists in interpretation, triaging \cite{Doi2007Computer-Aided}, and diagnostic decision-making \cite{Yu2011automatic}. In general, the process contains three steps: localize region of interest (ROI), extract features from the ROI, and apply machine learning models for the diagnosis or the detection of diseases \cite{Candemir2019review}. Significant research progress in deep learning has led to enhanced performance in analyzing CXR images. In \cite{Dunnmon2019Assessment}, the authors achieved an area under the curve (AUC) of $0.960$ on the classification of CXR images. Another research \cite{Tang2020Automated} on abnormality classification of CXR images led to an AUC of $0.982$. 

There are challenges in developing deep learning models for the analysis of CXR images. First, supervised deep learning relies on images with annotations; however, human-powered labeling for providing high-quality labels is labor-intensive, slow, and expensive. To alleviate this, natural language processing (NLP), as a fast and inexpensive alternative, is used to create image labels from medical records. Nevertheless, the NLP approach can be unreliable and produce label errors when applied alone. Additionally, some visible but minor abnormalities may have been ignored in the radiology reports. Therefore, this information will not be extracted by an NLP algorithm \cite{Olatunji2019Caveats,Irvin2019CheXpert,Call2021Deep} but can be essential for the diagnosis of diseases.

Secondly, there exists a large feature gap between pre-trained feature extractors and target data sets. Due to the limited number of training data, CXR image analysis normally relies on a transfer learning approach, which adopts a deep neural network pre-trained on a large natural image data set (e.g., ImageNet \cite{Deng2009ImageNet}), and then fine-tunes it on a target medical data set. This approach has demonstrated its effectiveness; however, the modalities of natural and medical images differ considerably, which can degrade transfer learning performance significantly \cite{Wen2021Rethinking,Raghu2019Transfusion}. 

\begin{figure}[!ht]
\centering
\includegraphics [width=0.9\textwidth]{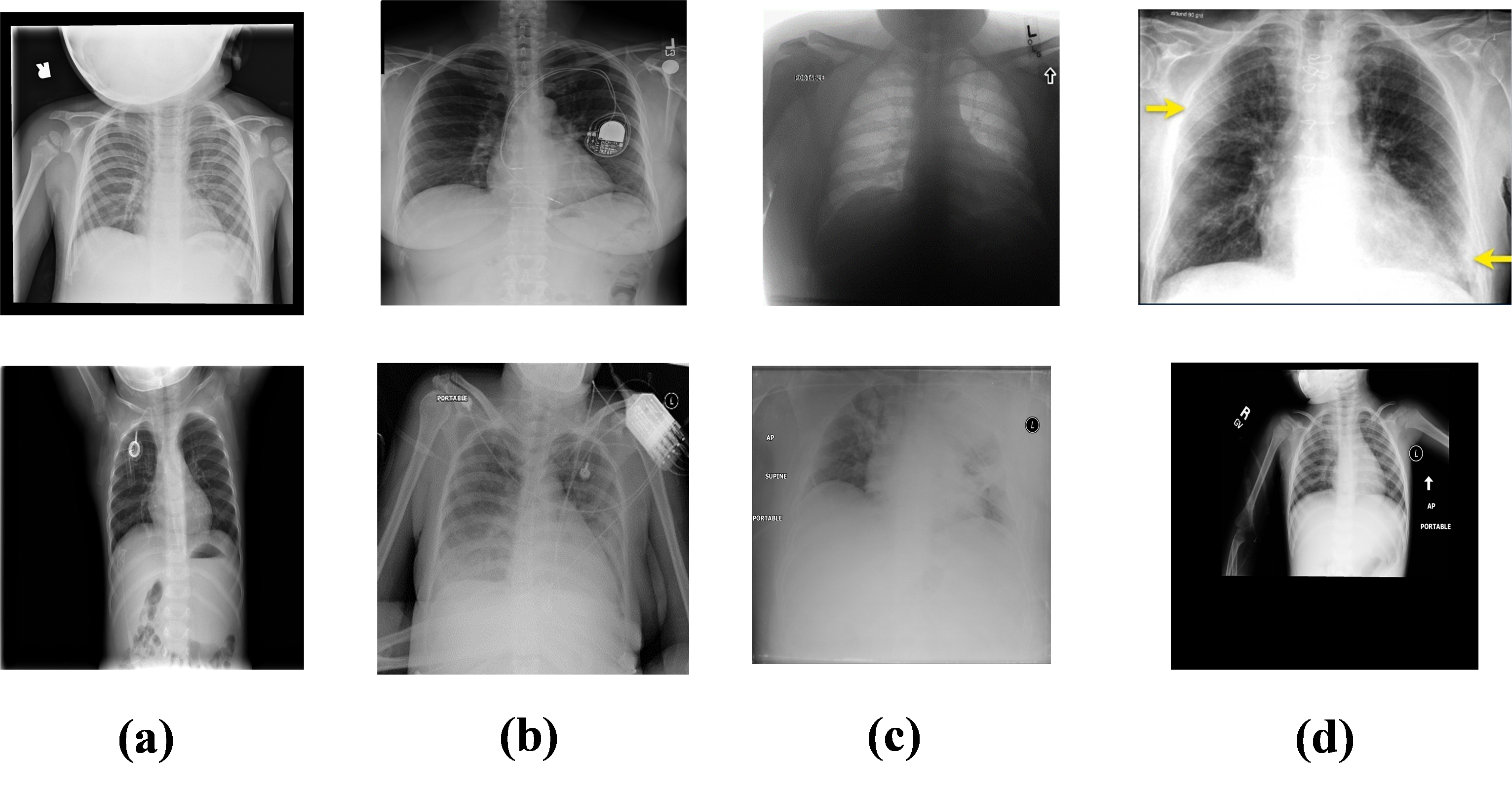}
\caption{Example of noisy and distorted CXR images from COVIDx dataset.}
\label{fig:1}
\end{figure}

Thirdly, characteristics associated with each CXR image data set also affect model performance. For example, COVIDx is an open-access benchmark data set and it comprises $16,352$ CXR images from a multinational cohort of $15,346$ patients from at least 51 countries \cite{Wang2020COVID-Net}. Like other CXR databases, the COVIDx data set contains a considerable number of images with peculiarities, some of which are demonstrated in Fig. \ref{fig:1}. Body parts, including a patient's head, hands, and abdomen, are often captured in pediatric CXR images (Fig. \ref{fig:1}-a). Foreign objects such as implanted devices and wires appear in CXR images (Fig. \ref{fig:1}-b). Some images are under-penetrated radiographs (Fig. \ref{fig:1}-c). In addition, there are cases where the lung is surrounded by a black box and/or with the presence of symbols and text (Fig. \ref{fig:1}-d). Moreover, parameters such as age, gender, heart dimension, and pathology cause lung appearance to vary among patients. These complexities can cause a computer-aided diagnostic system to make decisions based on irrelevant information. Therefore, we hypothesize that the localization and subsequent analysis of lung area can improve the performance of computer-aided diagnostic systems for CXR images.

To address the above challenges, we propose a deep learning approach to enhance the classification performance of CXR images. The approach is designed in a cascaded manner with two modules, a deep neural network with criss-cross attention modules (XLSor) for localizing lung region in CXR images, and a CXR classification model with a backbone of a self-supervised momentum contrast (MoCo) model pre-trained on large-scale CXR data sets. For the segmentation module, we compare two deep learning models, one inspired by U-Net (UNB) \cite{Ronneberger2015U-Net}, and the other comprising criss-cross attention modules (XLSor) \cite{Tang2019XLSor}. These two models are evaluated using both traditional measures, including Intersection-over-Union (IoU) \cite{Rezatofighi2019Generalized} and Dice coefficient \cite{Bertels2019Optimizing}, and by novel statistical metrics which we design specifically for CXR images without lung masking labels. Once the best model is chosen, it is used to generate bounding boxes as an automated pre-processing step for classification. We opt to use bounding boxes as opposed to fine lung segmentation masks to avoid information loss from potential segmentation errors. This leads to a more reliable and generalizable segmentation approach for CXR images. 

The extracted lung regions produced in the segmentation step are therefore used for training deep classification models. The COVIDx dataset  \cite{Wang2020COVID-Net} (Version 5) is used for the evaluation of the models. Since COVID-19 is not the focus of this study, Pneumonia and COVID-19 are combined and merged into one class called ``abnormal". As a result, deep learning models are trained and evaluated on a binary classification task (normal vs. abnormal). To train the classification models, we use Momentum Contrast (MoCo) \cite{Kaiming2019Moco} backbone pre-trained on large CXR data sets, and then fine-tune it on the segmented lung data set. As a self-supervised approach, the MoCo model was proposed to learn image representations using a large number of unlabelled data \cite{Kaiming2019Moco,Falcon2020Framework}. In this work, we adopt a MoCo model that has been pre-trained on large-scale CXR image benchmark databases \cite{Sriram2021CovidPrognosis}. Therefore, this backbone fits more closely with CXR classifications than the ImageNet data set. As a result, the proposed approach not only avoids the need for large-scale image labelings, but also addresses the feature gap.
 
Providing explainability for medical image classifiers is as important as achieving high performance. It is unacceptable to deploy models that do not use correct image features for making decisions, even if they have achieved high performance metrics. Therefore, we further evaluate the classifiers using Gradient-weighted Class Activation Mapping (Grad-CAM) \cite{Selvaraju2016gradCAM}. The Grad-CAM generates heat maps indicating regions most relevant for classification predictions \cite{8438639}. An advantage of Grad-CAM visualizations is its ability to be class-discriminative (\cite{Selvaraju2016gradCAM}. 

Our main contributions are as follows:
\begin{itemize}
  \item Proposed a cascaded design comprising segmentation and classification of CXR images for improved image classification with reliable features;
  \item Applied attention to the classification through selecting the best lung segmentation approach using novel statistical metrics for evaluation;
  \item Adopted a self-supervised deep learning model for the classification of CXR images, which addresses label needs and feature gaps;
  \item Studied model explainability through cross-checking different scenarios for the classification of CXR images with and without segmentation.
\end{itemize}

\section*{Methods}
\begin{figure}[!ht]
\centering
\includegraphics[width=0.99\linewidth]{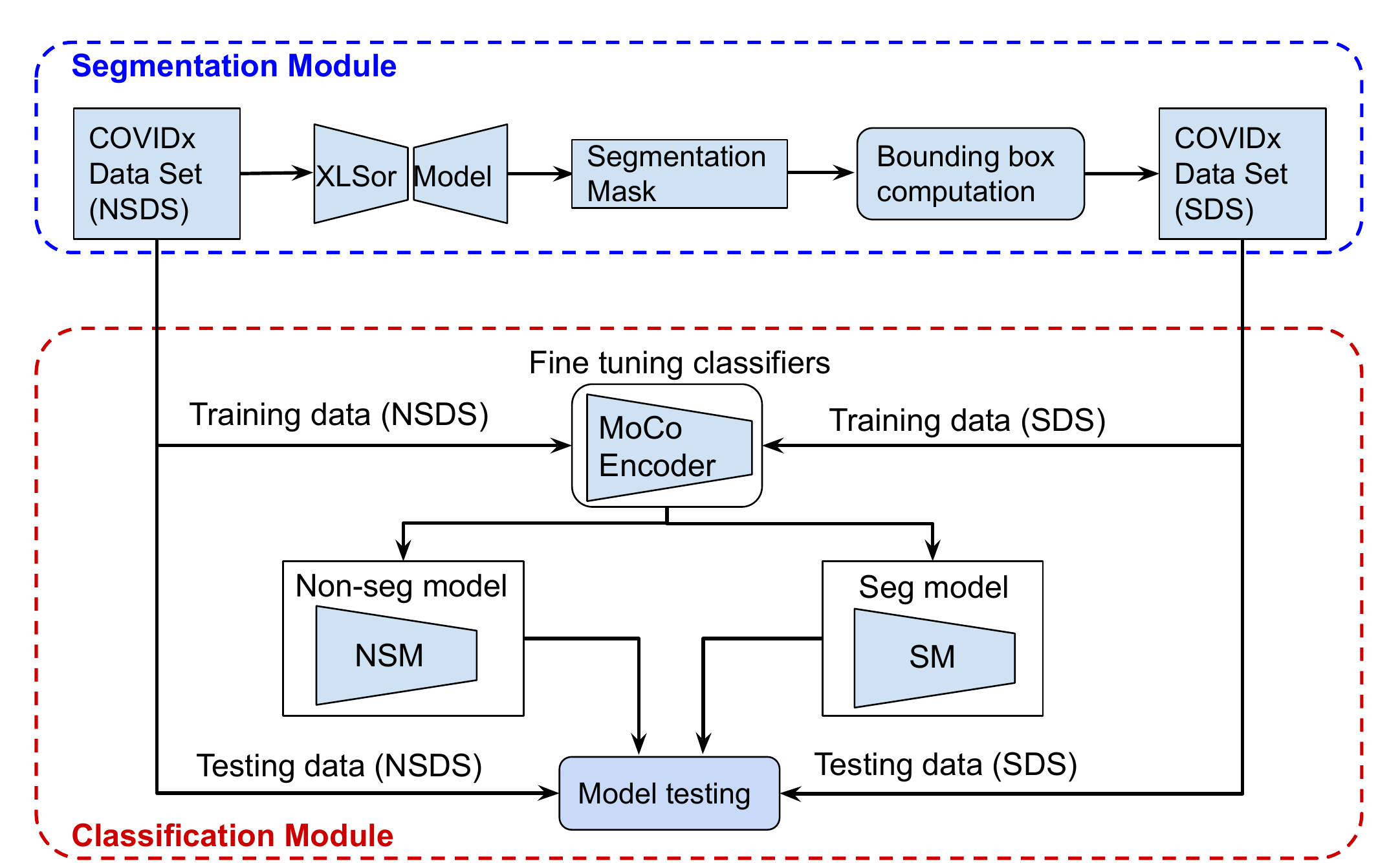}
\caption {The pipeline of proposed approach.}
\label{fig:2}
\end{figure}

Fig. \ref{fig:2} shows the pipeline of our proposed approach, which consists of two main modules: segmentation and classification. The non-segmented COVIDx data set (NSDS) is fed to the segmentation model to generate segmentation masks for the segmentation module. A bounding box is then calculated from the mask. Finally, the image is cropped based on the bounding box to generate the segmented data set (SDS). The SDS is then used to fine-tune a MoCo pre-trained encoder, generating the segmented model (SM). In parallel, NSDS, as the original non-segmented data set, is used to fine-tune a MoCo pre-trained encoder to generate a non-segmented model (NSM). Both the SDS and the NSDS are split into the same training and test splits, and test splits are reserved as unseen data for evaluation. The SM and NSM are evaluated on both the SDS test split and the NSDS test split. 
%The classification and segmentation approaches adopted and evaluated for our pipeline have different networks as backbones that desire different input sizes. Therefore, based on the backbone network for each step, CXR images must be resized to the size required by the network. Details of the pre-processing steps required for each module, including resizing, are provided comprehensively in the assigned subsections.

\subsection*{Dataset} 
The following two data sets are adopted in this paper:
\begin{itemize}
    \item Shenzhen Hospital (SH) Data Set: The SH data set contains 566 manually segmented lung masks from Shenzhen No.3 Hospital, Shenzhen, China. The dataset comprises normal and abnormal CXR images with markings of tuberculosis \cite{Stirenko2018Chest,Candemir2014Lung,Jaeger2014Automatic}. 
    \item COVIDx Data Set: The COVIDx Data Set consists of 15,279 CXR images from three different classes, Normal (8,851 images), Pneumonia (6,045 images), and COVID-19 (383 images). The data set is the largest open-access data set in terms of the number of Covid-19 positive patient cases \cite{Wang2020COVID-Net}. It should be noted that the COVIDx data set does not provide lung masks.
\end{itemize}

Both data sets are used for the evaluation of different segmentation approaches. In addition, the chosen segmentation approach will be applied to the COVIDx data set for generating SDS. Both versions of the COVIDx data set, SDS and NSDS, are used for fine-tuning and evaluating the classification models.

\subsection*{Model Architecture and Research Pipeline}
\subsubsection*{Segmentation}
To perform segmentation and extract lung areas from CXR images, we adopt and compare two different segmentation approaches, namely UNB \cite{Ronneberger2015U-Net} and XLSor \cite{Tang2019XLSor}. The UNB was explicitly invented to deal with biomedical images \cite{Ronneberger2015U-Net}, and it is very well-known in medical image segmentation \cite{AsgariTaghanaki2021Deep}. As an example, in \cite{Wang2017Segmentation}, a U-Net-based image segmentation outperformed most conventional methods on lung, heart, and clavicle segmentation in the chest radiograph, and it delivered better results on more challenging structures. U-Net comprises the encoder, which reduces the spatial dimension by continuously merging the layers to extract feature information, and the decoder, which restores the target detail and the spatial dimension according to the feature information \cite{Cai2020review}. Moreover, compared to other deep learning-based image segmentation, U-Net contains shortcut connections from the analysis path to expansion path between the layers of equal resolution, in order to provide essential high-resolution features to the deconvolution layers \cite{Hesamian2019Deep}.

The XLSor was invented to improve on the UNB. In \cite{Tang2019XLSor}, the authors claim that, when testing on unseen abnormal CXR images, the performance of a U-Net Based segmentation approach is not very promising due to capturing insufficient contextual information of the lungs. However, the XLSor can capture long-range contextual information using a fully convolutional network and two criss-cross attention modules with shared weights. The criss-cross attention module collects contextual information in horizontal and vertical directions to enhance pixel-wise representative capability \cite{Huang2020CCNet:}.

Both architectures were trained on Japanese Society of Radiological Technology (JSRT) \cite{Shiraishi2000Development}, and Montgomery County chest X-ray (MC) databases \cite{Jaeger2014Two}. In order to train a robust lung segmentation network on small data sets, each approach has a different strategy to enrich the training data. For UNB, a combination of affine transformations, such as shifting, zooming in/out, rotation and flipping, were applied to augmenting CXR images in real-time while the model was still in training. However, XLSor adopted MUNIT \cite{Huang2018Multimodal}, an image-to-image translation method (i.e., from normal CXR images to abnormal ones) for data augmentation. Moreover, UNB uses ADAM optimizer with cross-entropy loss function, while XLSor uses mean square error loss function and SGD optimizer.

\begin{figure}[!ht]
\centering
\includegraphics[width=0.98\linewidth]{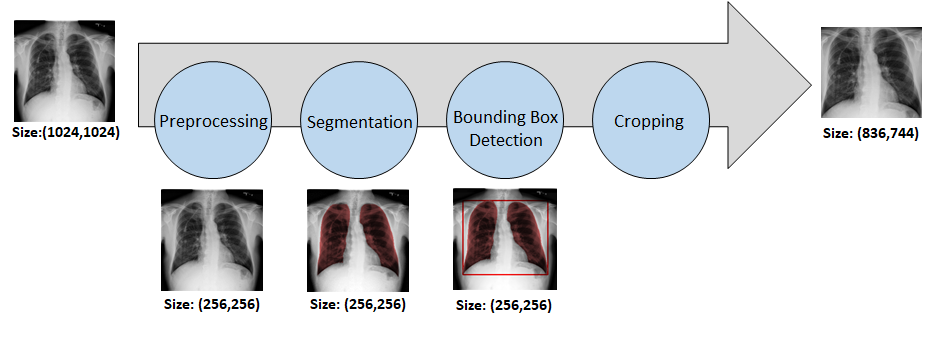}
\caption {Pipeline of segmentation and extracting lung area from CXR images.}
\label{fig:SegmentaionPipeline}
\end{figure} 

The segmentation pipeline is the same for both approaches, as illustrated in Fig. \ref{fig:SegmentaionPipeline}. First, the CXR images are resized. Then, the lung lobes are segmented by the trained models. Next, the bounding box is calculated based on the segmented areas. The coordinates of the bounding box are re-scaled based on the original size of the image. Finally, the original image is cropped using the coordinates of the re-scaled bounding box. As a result, the high-resolution images will be preserved for the classification step. We opt to use bounding boxes as opposed to the fine lung segmentation masks to avoid information loss from potential segmentation errors. This allows the segmentation module to be more reliable and generalized for CXR classification tasks. 

To evaluate both approaches, we first apply them to the SH data set, which contains lung masks. The models are also evaluated on the COVIDx data set. Since this data set does not contain lung masks, bounding boxes are computed based on segmentation outputs from the models. Then, statistical analysis on the centers of the bounding boxes (CBB) is conducted to compare model performance. As complementary parameters to CBB, two more parameters are defined as follows:
\begin{itemize}
    \item $SA/LLA$: Ratio of the smallest detected area (SA) to the largest detected area (LLA).
    \item $LA/LLA$: Ratio of the second-largest detected area (LA) to the largest detected area.
\end{itemize}

Contrary to CBB, these parameters do not depend on the location of the detected bounding box. Instead, they evaluate the proportion of two segmented lung lobes, i.e., left and right lobes.

\subsubsection*{Classification}

To perform classification tests, CXR images with uneven dimensions are padded. Next, the images are resized to $224\times224$. Finally, the pixel values of the images are normalized. After these pre-processing steps, we fine-tune a self-supervised model based on the MoCo method \cite{Kaiming2019Moco}, which has been pre-trained on MIMIC-CXR data set \cite{Johnson2019MIMIC-CXR} and CheXpert data set \cite{Irvin2019CheXpert}. Sriram et al. \cite{Sriram2021CovidPrognosis} provided these models developed in the pre-training step for COVID-19 prognosis. Despite performing a different classification task in this paper, we leverage the extractor that has been pre-trained on CXR data, as opposed to applying transfer learning on traditional models pre-trained on the ImageNet data set. Features extracted using the latter approach do not match those of medical data. This pre-trained MoCo model acts as the embedding extractor, and a linear classification layer with randomly initialized weights is appended afterward to complete the transfer learning architecture. The pre-trained model is then fine-tuned on the COVIDx data set.

We perform two different experiments to evaluate the effectiveness of our proposed pipeline. These experiments can be broadly classified into (1) Our proposed pipeline, i.e., classification of segmented lung regions from the COVIDx data set into normal or abnormal (2) Traditional classification implementation, i.e., classification of entire CXR images from the COVIDx data set into normal or abnormal. To distinguish the models, hereafter, we call them Segmented-Model (SM) for experiment (1) and Non-Segmented-Model (NSM) for experiment (2). For NSM, all segmentation blocks, including bounding box detection and image cropping, are not needed.

\begin{table}[t]
\caption{The train/test split of COVIDx-CXR data set}
\label{data-table}
\begin{center}
\begin{tabular}{lll}
\multicolumn{1}{c}{\bf COVIDx Split}  &\multicolumn{1}{c}{\bf Normal} &\multicolumn{1}{c}{\bf Abnormal}
\\ \hline \\
Train & 7966 & 5976 \\
Test & 885 & 694 \\
\end{tabular}
\end{center}
\end{table}

The process for training the SM and the NSM is as follows. The train/test split for the COVIDx data set is listed in Table \ref{data-table}. First, the Pneumonia and Covid-19 classes are combined into the abnormal class, creating a two-class (normal vs. abnormal) classification problem. We reserve the test split as unseen data until the end, and all classification results are evaluations on this test set. The extracted lung regions of the train split are used to train the SM, and full CXR images of the train split are used to train the NSM. Both train splits are further divided via a 0.1 validation split during training to track performance. Due to the uneven distribution of normal and abnormal classes, oversampling is used to sample all classes uniformly in each mini-batch. No data augmentation is used to isolate the characteristics of the segmented and non-segmented data sets during training. 

Training parameter settings are as follows. The SGD optimizer is used with a momentum of 0.9 and $1e-4$ weight decay. The learning rate is set at $1e-4$, and exponential step decay is applied as the learning rate reduction method during training. The step decay is applied at the same epochs for both models with steps of 4 and a decay factor of 0.94. Both models were trained for 100 epochs.

\subsection*{Performance assessment}
In this study, two types of deep learning models are involved in terms of their purposes: lung segmentation and CXR classification. For the segmentation models (i.e., UNB and XLSor), overlap-based metrics are adopted to evaluate their performances. These metrics quantify the overlapping area between reference boundary and predicted segmentation. The two most common metrics are: IoU \cite{Rezatofighi2019Generalized} and Dice coefficient \cite{Bertels2019Optimizing}. Both measures have a range of values between 0 and 1, where 1 indicates fully overlapped segmentation. For the COVIDx data set with no annotations of lung masks, instead of computing the overlap-based metrics, we design state-of-the-art parameters to compare structural information \cite{Santosh2018Automated,Karargyris2016Combination} from segmented lung areas and computed bounding boxes around the detected lung segments. These parameters have been introduced in the previous section.

For CXR classification models (i.e., both SM and NSM), after training and fine-tuning, each model is evaluated on the segmented variant of the test split, so-called SDS, and the non-segmented variant the test split, so-called NSDS. That is, the SM would be tested on both the SDS and NSDS and likewise for the NSM. This is done because CXR image quality can differ drastically in real-life applications, and a good model should perform well in both the segmented and non-segmented variants of CXRs. Evaluating the models on the test split from different experiments allows us to answer following questions:
\begin{itemize}
    \item Can providing extracted lung regions to train the SM model guide the model to focus on relevant information (i.e., lung area) in CXR images?
    \item How robust and generalized is the SM model when applied to challenging CXR images? 
    \item Considering that the NSM is trained on challenging and noisy images,  would it have better performance on SDS as simplified and de-noised images, comparing to its performance on NSDS?
\end{itemize}

On top of numerical metrics (accuracy, positive predictive value, sensitivity, F1 score), we also adopt Grad-CAM \cite{Selvaraju2016gradCAM} to compute a visual representation of how the models make predictions. For medical image analysis, it is imperative to have confidence in knowing that the models are making decisions based on the correct information. To achieve this, Grad-CAM \cite{Selvaraju2016gradCAM} is applied to the final convolution layer in the models to produce a coarse localization map highlighting important regions on a CXR image that each model uses to make predictions. 

\section*{Results and Discussion}
We evaluate the segmentation and classification modules through 1) adopting novel evaluation metrics and methods, 2) testing on different data sets, 3) fine-tuning the networks, and 4) visualization by Grad-CAM. In the following, we will discuss and interpret all the results obtained through these evaluation processes.

\subsection*{Segmentation}
We nominated two different deep learning-based segmentation approaches pre-trained on CXR data sets. To select one of the methods for the next step of our designed pipeline, we first applied them to the SH data set containing lung masks. Table \ref{table1} indicates the performance of the two segmentation approaches on the SH data set based on Dice and IoU metrics averaged over all 566 images. XLSor outperformed UNB with a mean IoU of 0.938 and a mean Dice of 0.968. Our results are consistent with what is stated in \cite{Tang2019XLSor}, where U-Net \cite{Ronneberger2015U-Net} was applied for comparison to demonstrate the effectiveness of the criss-cross attention-based XLSor. In \cite{Tang2019XLSor}, XLSor also achieved better results than U-Net on different data sets.
\begin{table}[h!]
\caption{Performance of the two segmentation approaches on the SH data set.}
\label{table1}
\begin{center}
\begin{tabular}{lll}
\multicolumn{1}{c}{\bf}&\multicolumn{1}{c}{\bf Mean IoU}&\multicolumn{1}{c}{\bf Mean Dice} 
\\ \hline \\
UNB & 0.913 & 0.954 \\
XLSor & 0.938 & 0.968 \\
\end{tabular}
\end{center}
\end{table}

Challenges such as deformation due to pulmonary diseases may cause the lung region not to be accurately and entirely annotated by a segmentation algorithm. Indeed, the missing lung areas from the annotation are mostly those that contain essential information about an abnormality. The presence of these areas is necessary for further processing, especially for computer-aided diagnosis. Considering the good performance of both networks on the SH data set, it is expected that they would have a robust performance on the majority of COVIDx images. 

\begin{figure}%
\centering
\subfigure[]{%
\includegraphics[height=1.7in]{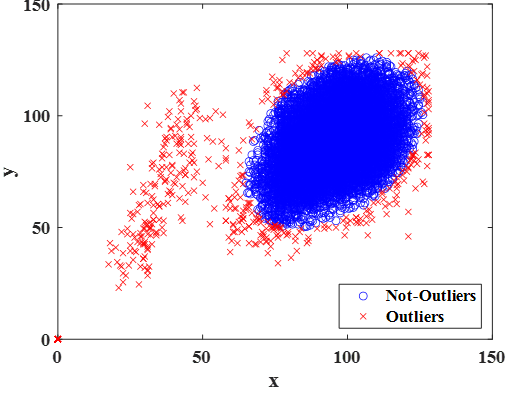}}%
\qquad
\subfigure[]{%
\includegraphics[height=1.7in]{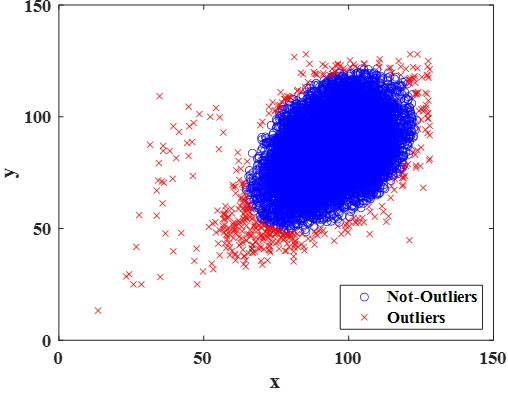}}%
\caption{Centers of bounding boxes computed using (a) UNB and (b) XLSor.}
\label{fig:3}
\end{figure}

We further evaluate the two segmentation models on the COVIDx data set. Fig. \ref{fig:3} illustrates CBBs, calculated for each CXR image separately. Fast minimum covariance determinant  (FAST-MCD) \cite{Rousseeuw1999Fast} was applied to CBBs of all COVIDx images to find the outliers. Here, outliers are defined as those CXR images in which the extracted information from their segmented lung areas and detected bounding boxes have different behavior from most images. The FAST-MCD method selects $h$ observations out of $n$ (where $n/2 < h \leq n$) whose classical covariance matrix has the lowest possible determinant. The MCD mean is the mean of the $h$ selected observations. 

According to Fig. \ref{fig:3}, it can be seen that the majority of CBBs (in blue) in both methods are placed in the same area of CXR images (i.e., same lung area in CXR images). The main difference is related to the outliers (CBBs in red). In UNB, many outliers are far from the blue mass (see Fig. \ref{fig:3}-a). Additionally, in contrast to XLSor, UNB could not detect the lung area and therefore failed to compute bounding boxes for 30 images (i.e., outliers located at the origin of the coordinate system in Fig. \ref{fig:3}-a labeled with red cross). These images were mostly cases surrounded by large black frames (Fig. \ref{fig:1}-d) or of under-penetrated radiographs (Fig. \ref{fig:1}-c). However, as the outputs of XLSor, the outliers are mostly placed around the blue mass with no samples at the origin of the coordinate (see Fig. \ref{fig:3}-b). 

\begin{table}[ht!]
\caption{Number of outliers per class of COVIDx dataset as a result of FAST-MCD.}
\label{table2}
\begin{center}
\begin{tabular}{llll}
\multicolumn{1}{c}{\bf } & \multicolumn{1}{c}{\bf Abnormal}&\multicolumn{1}{c}{\bf Normal}&\multicolumn{1}{c}{\bf Total} 
\\ \hline \\
UNB & 374 (5.8\%) & 120 (1.4\%) & 494 (3.2\%) \\
XLSor & 283 (4.4\%) & 146 (1.6\%) & 429 (2.8\%) \\
\end{tabular}
\end{center}
\end{table}

For quantitative comparison, Table \ref{table2} lists the number of outliers for both classes of the data set as the output of FAST-MCD. Also, the results were calculated in percentage by considering the number of observations in each class. Applying the XLSor, slightly more outliers were detected than applying the UNB to the class of ``Normal" (1.6\%). However, the overall performance of XLSor (2.8\%) is better than UNB (3.2\%). It should be considered that the blue mass for XLSor (Fig. \ref{fig:3}-b) is denser than UNB's (Fig. \ref{fig:3}-a). Given that the outliers are defined based on blue masses containing the majority of samples' CBBs with the same behavior, those samples counted as outliers for XLSor may not be detected as outliers for UNB. Indeed the evaluation process is more strict for XLSor than UNB, and this still confirms the better performance of XLSor.

\begin{figure}%
\centering
\subfigure[]{%
\includegraphics[height=1.5in]{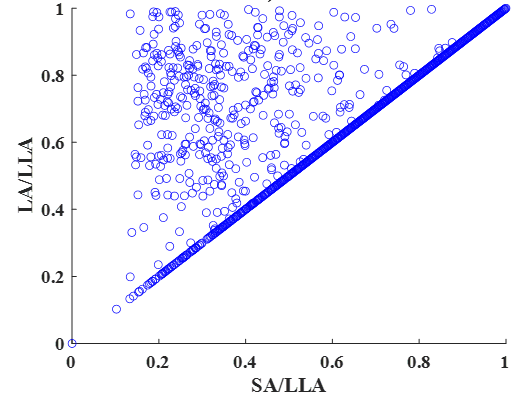}}%
\qquad
\subfigure[]{%
\includegraphics[height=1.5in]{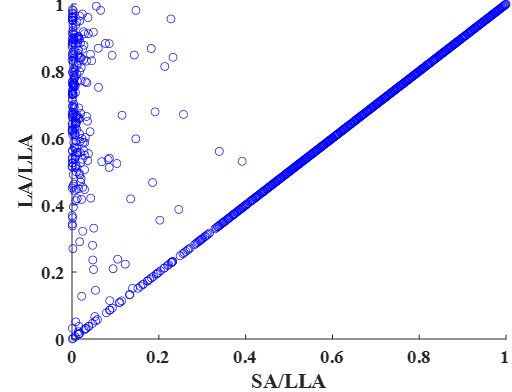}}%
\caption{(LA/LLA) as a function of (SA/LLA). (a) UNB, (b) XLSor.}
\label{fig:4}
\end{figure}

\begin{figure}%
\centering
\subfigure[]{%
\includegraphics[height=1.7in]{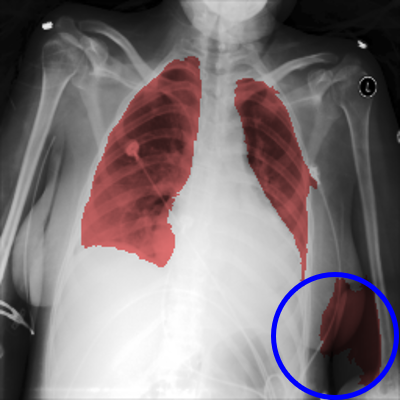}}%
\qquad
\subfigure[]{%
\includegraphics[height=1.7in]{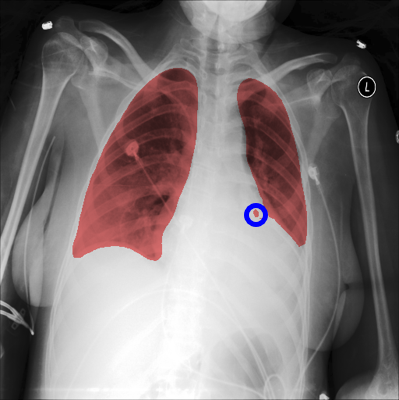}}%
\caption{Example of detected lung region by the two networks in the same image (a: UNB, b: XLSor).}
\label{fig:5}
\end{figure}

When applying a segmentation network, multiple regions may be returned as detected lung areas from each image. To complement the CBB, SA/LLA (Ratio  of  the  smallest  detected  area to  the  largest  detected area) and LA/LLA (Ratio of the second-largest detected area to the largest detected area) are used to evaluate lung lobe detection. Ideally, a reliable automated lung segmentation approach must find two complete lung lobes (i.e., two regions). In this case, (LA/LLA) and (SA/LLA) will be identical. Fig. \ref{fig:4} illustrates (LA/LLA) as a function of (SA/LLA). Commonly for both methods, the majority of observations follow an identity function, which means the algorithms mostly found two lung lobes per image. For those observations that did not follow the identity function, the two methods behaved differently. The difference is mainly rooted in two scenarios. Fig. \ref{fig:5} shows an example of a detected lung region by the two networks in the same image. In challenging CXR images, the methods may find more than one region per lung lobe (XLSor, Fig. \ref{fig:5}-b). In this scenario, the segmented area does not have a significant effect on the final boundary box. However, in the second scenario, the algorithm mistakenly selects another area excluding the lung lobes (e.g., other parts of the body or background) as part of the lung area (UNB, Fig. \ref{fig:5}-a). In this scenario, the final bounding box may cover a larger area than the lung region. Although the segmentation approach would not cause any loss of information for the second scenario, it questions the main purpose of lung segmentation (i.e., removing irrelevant regions and accurate localization of the lung area). As illustrated in Fig. \ref{fig:4}, while XLSor mostly followed the first scenario, UNB followed the second one. Therefore, XLSor is a better segmentation network than UNB.

\begin{table}[ht!]
\caption{Number of outliers from the COVIDx dataset as a result of computing ratio parameters: (LA/LLA) and (SA/LLA).}
\label{table3}
\begin{center}
\begin{tabular}{llll}
\multicolumn{1}{c}{\bf } & \multicolumn{1}{c}{\bf Abnormal}&\multicolumn{1}{c}{\bf Normal}&\multicolumn{1}{c}{\bf Total} 
\\ \hline \\
UNB  & 343 (5.3\%)  & 66 (0.7\%)  & 409 (2.7\%) \\
XLSor  & 219 (3.4\%)  & 47 (0.5\%)  & 266 (1.7\%) \\
\end{tabular}
\end{center}
\end{table}

For each method, Table \ref{table3} lists the number of observations which did not follow the identity function in Fig. \ref{fig:4}. Overall, the XLSor has better performance than the UNB with fewer outliers. 

Considering the results and our observation, we chose XLSor as the segmentation approach for our pipeline. Therefore, the extracted lung regions from the COVIDx data set as the output of the XLSor were the input of the classification step. 
Despite our comprehensive evaluation of the two different segmentation approaches, it is noted that the ultimate goal is not segmentation optimization in this paper but rather to investigate its actual benefit on the resulting classification.

\subsection*{Classification}
\begin{table}[t]
\caption{Performance of MoCo models on different test data set}
\label{classification-result}
\begin{center}
\begin{tabular}{lllll}
\multicolumn{1}{c}{\bf Model $\rightarrow$ Dataset}  &\multicolumn{1}{c}{\bf Accuracy} &\multicolumn{1}{c}{\bf Sensitivity} &\multicolumn{1}{c}{\bf Precision} &\multicolumn{1}{c}{\bf F1-Score}
\\ \hline \\
NSM $\rightarrow$ NSDS & 0.946 & 0.935 & 0.942 & 0.939\\
NSM $\rightarrow$ SDS & \textbf{0.870} & 0.973 & \textbf{0.783} & 0.868\\
SM $\rightarrow$ SDS & 0.946 & 0.931 & 0.944 & 0.938\\
SM $\rightarrow$ NSDS & 0.928 & 0.860 & 0.974 & 0.914\\
\end{tabular}
\end{center}
\end{table}

\begin{figure}[!ht]
\centering
\includegraphics[width=0.99\linewidth]{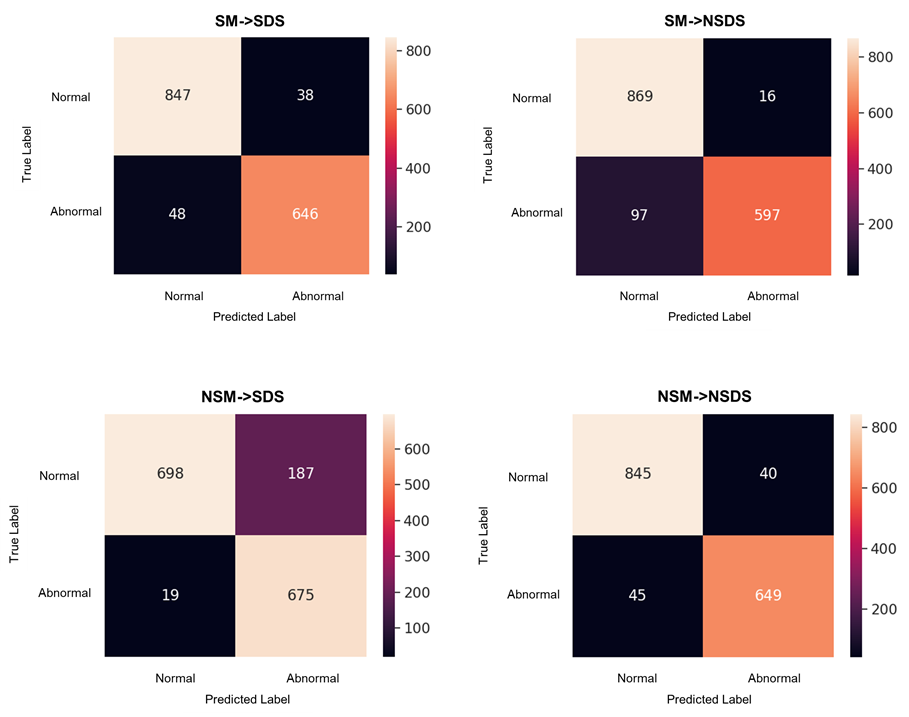}
\caption {Confusion matrices of applying different models to different test data sets.}
\label{fig:Conf_Metrix}
\end{figure}

As discussed in the previous section, we optimized two models for the classification of CXR images: SM and NSM. For a comprehensive evaluation, these models were applied on both SDS data containing extracted lung areas and NSDS data containing entire CXR images. Table \ref{classification-result} lists the results of evaluating the models on different test data sets. The ($\rightarrow$) symbol represents applying a specific model to a specific test data set. The evaluation metrics for different scenarios have been calculated using the confusion matrices shown in Fig. \ref{fig:Conf_Metrix}. 

For SM tested on SDS and NSM tested on NSDS, both models achieved an accuracy of 0.946. In other words, when the models were tested on the data variant they were trained on, their performance was the same. When the SM was applied on the NSDS, there was a slight drop in accuracy from 0.946 to 0.928 (1.8\%), along with a sensitivity drop from 0.931 to 0.860 (i.e., more false-negative cases). However, when the NSM was applied to SDS, a more severe accuracy drop of 0.946 to 0.870 (7.6\%) with a corresponding sharp precision drop of 0.942 to 0.783 (i.e., more false-positive cases) was observed. Therefore comparatively, the SM performed much better on the NSDS than the NSM on the SDS. While the two models have similar performance on the data variants they were trained on, the cross-check results show that the NSM generalizes poorly on the SDS, whereas the SM still retrains good performance on the NSDS. Since the SDS contains only lung regions, the NSM's significant performance degradation suggests its use of clinically irrelevant regions to make predictions. We verify this hypothesis with Grad-CAM.

\begin{figure}[!ht]
\centering
\includegraphics[width=0.75\linewidth]{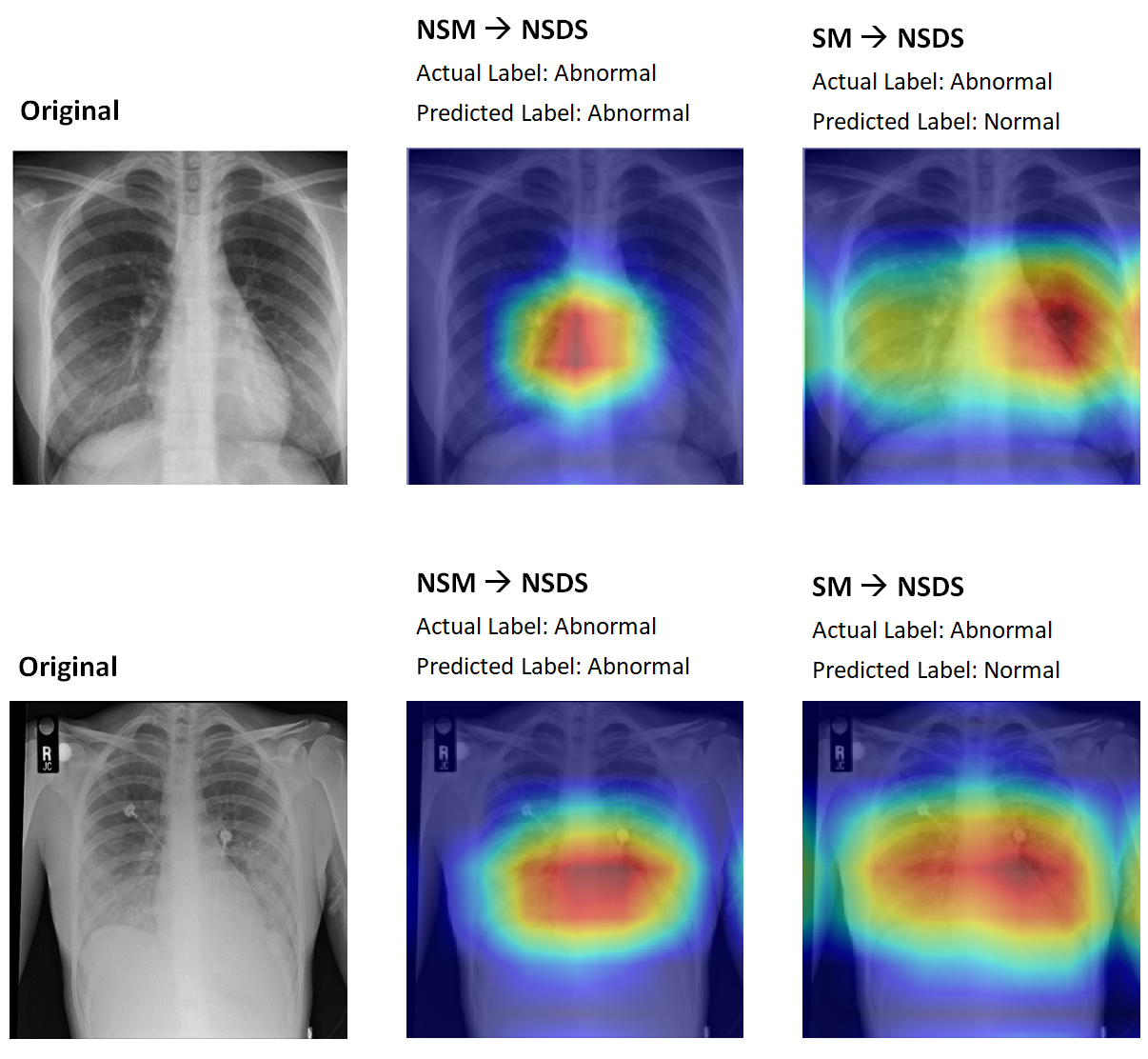}
\caption {Grad-CAM visualization of SM and NSM performances on NSDS.}
\label{fig:Grad_CAM_NSM}
\end{figure}

To explain the performance differences observed in Table \ref{classification-result}, we computed Grad-CAM heat maps and selected those that contain the general trend of model attention when applied on the two different test splits. Fig. \ref{fig:Grad_CAM_NSM} contains examples where NSM classified the abnormal NSDS case correctly, but SM did not. The first image is the original image from left to right in each row, followed by a heat map generated from Grad-CAM for NSM applied to NSDS and SM applied to NSDS. Red areas in heat maps refer to the regions the network deems important for prediction. In both examples, although the NSM made the correct prediction, it focused on the spinal area, mistaking it as a lung abnormality.

On the other hand, the SM focused on the lung region to decide, despite being wrong. This highlights why a medical image classifier should not be chosen based exclusively on numerical metrics. In addition, the features the classifier uses to make predictions should also be taken into consideration. Although the NSM seems to be a classifier as good as SM in terms of its accuracy, it failed to focus on clinically meaningful abnormal regions of the CXR images to detect abnormal cases. Instead, it predicted abnormality based on the wrong features, making it undesirable as a practical classifier.

\begin{figure}[!ht]
\centering
\includegraphics[width=0.97\linewidth]{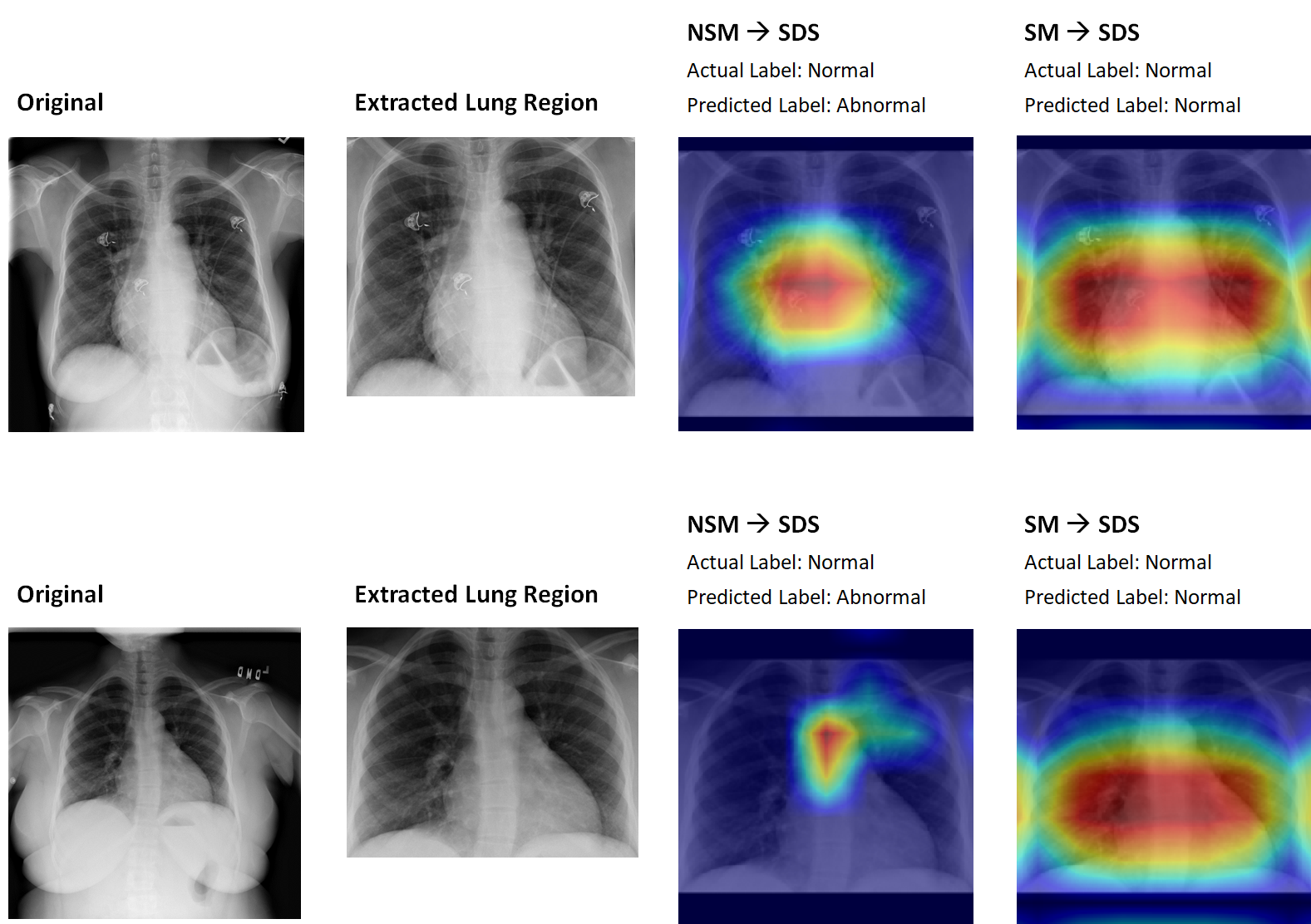}
\caption {Grad-CAM visualization of SM and NSM performances on SDS.}
\label{fig:Grad_CAM_SM}
\end{figure}

Fig. \ref{fig:Grad_CAM_SM} contains examples that demonstrate a complementary trend where the SM classified the normal SDS case correctly, but NSM did not. From left to right in each row, the first image is the original image; the second image is the extracted lung region as the output of XLSor, followed by heat map generated from Grad-CAM for NSM applied to SDS and finally, SM applied to SDS. As in Fig. \ref{fig:Grad_CAM_NSM} with the NSDS, the SM focuses correctly on the lung regions for SDS in Fig. \ref{fig:Grad_CAM_SM}. Comparatively, the NSM heat maps focus on irrelevant whiteness on CXR images such as the sternum (i.e., breastbone) or spine instead of the lungs, causing the classifier to over-classify abnormal cases. The model falsely predicts these regions as parts of the lung affected by COVID-19 or Pneumonia. One possible reason is that symptoms of Pneumonia or COVID-19, including lung inflammation, glass opacity, peripheral linear opacity, and consolidation, are also seen as whiteness in the lungs on CXR images depending on the severity of the disease. In addition, white shadows on CXR can represent dense or solid tissues, such as bones or the heart. In contrast, in normal CXR images, healthy lungs that predominantly contain air appear black \cite{Cleverley2020role}. This issue can clearly be very problematic, especially since the color spectrum of x-ray images is limited to gray-scale for different entities. The lung regions in a CXR image typically occupy around one-third or even less of the entire image \cite{Liu2019SDFN}. Thus, a model needs to first properly recognize the lung area and only then begin looking for whiteness to be effective amidst these distractions. The NSM evidently has not learned to do this and is biased toward any whiteness detected in the image as a sign of lung abnormality. On the other hand, the SM, which was trained on extracted lung images, seems to focus on the lung region for NSDS as well as SDS. The model trained with the segmented data set (SDS) was able to distinguish whiteness in the lung region specifically, with Grad-CAM heat maps concentrated on clinically relevant regions for both NSDS and SDS. This made it generalize well for both data sets and reduced performance degradation significantly.

If the model evaluation was based solely on accuracy and on corresponding test split (NSM on NSDS, SM on SDS), the advantage of applying the segmentation pre-processing step is not apparent; however, the significant performance degradation of the NSM on SDS and the Grad-CAM heat maps proved the opposite. The proposed segmentation step helped the classifier learn to focus more on clinically meaningful regions of the CXR images. Moreover, the cross-check results revealed that SM has better performance than the NSM model when facing a different set of data other than the one trained on. Therefore, in our proposed pipeline, we choose SM over NSM due to the improved reliability and generalizability of the model.

\section*{Conclusion}
In this study, we developed a deep learning-based approach to distinguish abnormal CXR images from normal ones. Validations of the approach suggest that applying automated segmentation as a pre-processing step for classification improves the generalization capability and the performance of the classification models. A distinctive characteristic of our approach is the use of features from the most meaningful part (i.e., the lung regions) of the CXR images to improve classification accuracy and reliability. In particular, we demonstrate that irrelevant features extracted outside the lung regions have misled the classifier's decision. In a worse scenario, the classifier focused on meaningless features despite making a correct decision. 

Our investigation suggests that segmentation has a considerable impact on CXR image classification in extracting and using reliable features, and it should be applied as a pre-processing step for the classification of CXR images. The adoption of a self-supervised learning approach also leads to excellent model performance while addressing the many challenges in medical image analysis.

The future scope of this study is to go one step further than CXR binary classification and evaluate our proposed approach facing different types of lung diseases, including COVID-19. Given the excellent performance of the MoCo model in this study, we plan to conduct further analysis on the different self-supervised approaches (e.g., SimCLR \cite{chen2020simple}) to analyzing CXR images.

% \section*{Appendix}
% Text for this section\ldots

%%%%%%%%%%%%%%%%%%%%%%%%%%%%%%%%%%%%%%%%%%%%%%
%%                                          %%
%% Backmatter begins here                   %%
%%                                          %%
%%%%%%%%%%%%%%%%%%%%%%%%%%%%%%%%%%%%%%%%%%%%%%
\subsubsection*{Acknowledgements}%% if any
Not applicable.

\subsubsection*{Funding}%% if any
This project is supported by Pandemic Response Challenge Program at the National Research Council Canada (NRC).

\subsubsection*{Abbreviations}%% if any
AUC: Area Under The Curve; CXR: Chest X-Ray; FAST-MCD: Fast-Minimum Covariance Determinant; Grad-CAM: Gradient-weighted Class Activation Mapping; IoU: Intersection-Over-Union; JSRT: Japanese Society of Radiological Technology; LA: Second-Largest Detected Area; LLA: Largest Detected Area; MC: Montgomery County; MoCo: Momentum Contrast; NLP: Natural Language Processing; NSDS: Non-Segmented Data Set; NSM: Non-Segmented Model; SA: Smallest Detected Area (SA); SDS: Segmented Data Set; SH: Shenzhen; SM: Segmented Model; UNB: U-Net Based; XLSor: X-ray Lung Segmentor.

\subsubsection*{Availability of data and materials}%% if any
The data sets used in this paper are publicly available and can be accessed from: 
\begin{itemize}
    \item \text{COVIDx}: \href{url}{https://github.com/lindawangg/COVID-Net/blob/master/docs/COVIDx.md}
    \item \text{SH}: \href{url}{https://www.kaggle.com/yoctoman/shcxr-lung-mask}
\end{itemize}

\subsubsection*{Ethics approval and consent to participate}%% if any
Not applicable.

\subsubsection*{Competing interests}
The authors declare that they have no competing interests.

\subsubsection*{Consent for publication}%% if any
Not applicable.

\subsubsection*{Authors' contributions}
Hilda Azimi \& Jianxing Zhang: Data Processing, Pipeline Design, Technical Validation, Visualization, Quality Control. Pengcheng Xi: Conceptualization, Pipeline Design, Technical Validation, Visualization, Quality Control. Hala As’ad: Quality Control. Ashkan Ebadi: Quality Control. St\'ephane Tremblay: Project Administration, Supervision. Alexander Wong: Conceptualization, Supervision. All authors read and approved the final manuscript.

\subsubsection*{Authors' information}%% if any
Not applicable.
%%%%%%%%%%%%%%%%%%%%%%%%%%%%%%%%%%%%%%%%%%%%%%%%%%%%%%%%%%%%%
% if your bibliography is in bibtex format, use those commands:
\bibliographystyle{bmc-mathphys} % Style BST file (bmc-mathphys, vancouver, spbasic).
\bibliography{bmc_article}      % Bibliography file (usually '*.bib' )


\begin{thebibliography}{39}
\ifx \bisbn   \undefined \def \bisbn  #1{ISBN #1}\fi
\ifx \binits  \undefined \def \binits#1{#1}\fi
\ifx \bauthor  \undefined \def \bauthor#1{#1}\fi
\ifx \batitle  \undefined \def \batitle#1{#1}\fi
\ifx \bjtitle  \undefined \def \bjtitle#1{#1}\fi
\ifx \bvolume  \undefined \def \bvolume#1{\textbf{#1}}\fi
\ifx \byear  \undefined \def \byear#1{#1}\fi
\ifx \bissue  \undefined \def \bissue#1{#1}\fi
\ifx \bfpage  \undefined \def \bfpage#1{#1}\fi
\ifx \blpage  \undefined \def \blpage #1{#1}\fi
\ifx \burl  \undefined \def \burl#1{\textsf{#1}}\fi
\ifx \doiurl  \undefined \def \doiurl#1{\textsf{#1}}\fi
\ifx \betal  \undefined \def \betal{\textit{et al.}}\fi
\ifx \binstitute  \undefined \def \binstitute#1{#1}\fi
\ifx \binstitutionaled  \undefined \def \binstitutionaled#1{#1}\fi
\ifx \bctitle  \undefined \def \bctitle#1{#1}\fi
\ifx \beditor  \undefined \def \beditor#1{#1}\fi
\ifx \bpublisher  \undefined \def \bpublisher#1{#1}\fi
\ifx \bbtitle  \undefined \def \bbtitle#1{#1}\fi
\ifx \bedition  \undefined \def \bedition#1{#1}\fi
\ifx \bseriesno  \undefined \def \bseriesno#1{#1}\fi
\ifx \blocation  \undefined \def \blocation#1{#1}\fi
\ifx \bsertitle  \undefined \def \bsertitle#1{#1}\fi
\ifx \bsnm \undefined \def \bsnm#1{#1}\fi
\ifx \bsuffix \undefined \def \bsuffix#1{#1}\fi
\ifx \bparticle \undefined \def \bparticle#1{#1}\fi
\ifx \barticle \undefined \def \barticle#1{#1}\fi
\ifx \bconfdate \undefined \def \bconfdate #1{#1}\fi
\ifx \botherref \undefined \def \botherref #1{#1}\fi
\ifx \url \undefined \def \url#1{\textsf{#1}}\fi
\ifx \bchapter \undefined \def \bchapter#1{#1}\fi
\ifx \bbook \undefined \def \bbook#1{#1}\fi
\ifx \bcomment \undefined \def \bcomment#1{#1}\fi
\ifx \oauthor \undefined \def \oauthor#1{#1}\fi
\ifx \citeauthoryear \undefined \def \citeauthoryear#1{#1}\fi
\ifx \endbibitem  \undefined \def \endbibitem {}\fi
\ifx \bconflocation  \undefined \def \bconflocation#1{#1}\fi
\ifx \arxivurl  \undefined \def \arxivurl#1{\textsf{#1}}\fi
\csname PreBibitemsHook\endcsname

%%% 1
\bibitem{Doi2007Computer-Aided}
\begin{barticle}
\bauthor{\bsnm{Doi}, \binits{K.}}:
\batitle{Computer-aided diagnosis in medical imaging: Historical review,
  current status and future potential}.
\bjtitle{Computerized medical imaging and graphics : the official journal of
  the Computerized Medical Imaging Society}
\bvolume{31}(\bissue{4-5}),
\bfpage{198}--\blpage{211}
(\byear{2007}).
doi:\doiurl{10.1016/j.compmedimag.2007.02.002}
\end{barticle}
\endbibitem

%%% 2
\bibitem{Yu2011automatic}
\begin{barticle}
\bauthor{\bsnm{Yu}, \binits{P.}},
\bauthor{\bsnm{Xu}, \binits{H.}},
\bauthor{\bsnm{Zhu}, \binits{Y.}},
\bauthor{\bsnm{Yang}, \binits{C.}},
\bauthor{\bsnm{Sun}, \binits{X.}},
\bauthor{\bsnm{Zhao}, \binits{J.}}:
\batitle{An automatic computer-aided detection scheme for pneumoconiosis on
  digital chest radiographs}.
\bjtitle{Journal of Digital Imaging}
\bvolume{24}(\bissue{3}),
\bfpage{382}--\blpage{393}
(\byear{2011}).
doi:\doiurl{10.1007/s10278-010-9276-7}.
\bcomment{PMID: 20174852 PMCID: PMC3092047}
\end{barticle}
\endbibitem

%%% 3
\bibitem{Candemir2019review}
\begin{barticle}
\bauthor{\bsnm{Candemir}, \binits{S.}},
\bauthor{\bsnm{Antani}, \binits{S.}}:
\batitle{A review on lung boundary detection in chest x-rays}.
\bjtitle{International Journal of Computer Assisted Radiology and Surgery}
\bvolume{14}(\bissue{4}),
\bfpage{563}--\blpage{576}
(\byear{2019}).
doi:\doiurl{10.1007/s11548-019-01917-1}
\end{barticle}
\endbibitem

%%% 4
\bibitem{Dunnmon2019Assessment}
\begin{barticle}
\bauthor{\bsnm{Dunnmon}, \binits{J.A.}},
\bauthor{\bsnm{Yi}, \binits{D.}},
\bauthor{\bsnm{Langlotz}, \binits{C.P.}},
\bauthor{\bsnm{Ré}, \binits{C.}},
\bauthor{\bsnm{Rubin}, \binits{D.L.}},
\bauthor{\bsnm{Lungren}, \binits{M.P.}}:
\batitle{Assessment of convolutional neural networks for automated
  classification of chest radiographs}.
\bjtitle{Radiology}
\bvolume{290}(\bissue{2}),
\bfpage{537}--\blpage{544}
(\byear{2019}).
doi:\doiurl{10.1148/radiol.2018181422}.
\bcomment{publisher: Radiological Society of North America}
\end{barticle}
\endbibitem

%%% 5
\bibitem{Tang2020Automated}
\begin{barticle}
\bauthor{\bsnm{Tang}, \binits{Y.-X.}},
\bauthor{\bsnm{Tang}, \binits{Y.-B.}},
\bauthor{\bsnm{Peng}, \binits{Y.}},
\bauthor{\bsnm{Yan}, \binits{K.}},
\bauthor{\bsnm{Bagheri}, \binits{M.}},
\bauthor{\bsnm{Redd}, \binits{B.A.}},
\bauthor{\bsnm{Brandon}, \binits{C.J.}},
\bauthor{\bsnm{Lu}, \binits{Z.}},
\bauthor{\bsnm{Han}, \binits{M.}},
\bauthor{\bsnm{Xiao}, \binits{J.}},
\bauthor{\bsnm{Summers}, \binits{R.M.}}:
\batitle{Automated abnormality classification of chest radiographs using deep
  convolutional neural networks}.
\bjtitle{npj Digital Medicine}
\bvolume{3}(\bissue{1}),
\bfpage{70}
(\byear{2020}).
doi:\doiurl{10.1038/s41746-020-0273-z}
\end{barticle}
\endbibitem

%%% 6
\bibitem{Olatunji2019Caveats}
\begin{botherref}
\oauthor{\bsnm{Olatunji}, \binits{T.}},
\oauthor{\bsnm{Yao}, \binits{L.}},
\oauthor{\bsnm{Covington}, \binits{B.}},
\oauthor{\bsnm{Rhodes}, \binits{A.}},
\oauthor{\bsnm{Upton}, \binits{A.}}:
Caveats in generating medical imaging labels from radiology reports.
arXiv:1905.02283 [cs, eess]
(2019).
arXiv: 1905.02283
\end{botherref}
\endbibitem

%%% 7
\bibitem{Irvin2019CheXpert}
\begin{botherref}
\oauthor{\bsnm{Irvin}, \binits{J.}},
\oauthor{\bsnm{Rajpurkar}, \binits{P.}},
\oauthor{\bsnm{Ko}, \binits{M.}},
\oauthor{\bsnm{Yu}, \binits{Y.}},
\oauthor{\bsnm{Ciurea{-}Ilcus}, \binits{S.}},
\oauthor{\bsnm{Chute}, \binits{C.}},
\oauthor{\bsnm{Marklund}, \binits{H.}},
\oauthor{\bsnm{Haghgoo}, \binits{B.}},
\oauthor{\bsnm{Ball}, \binits{R.L.}},
\oauthor{\bsnm{Shpanskaya}, \binits{K.S.}},
\oauthor{\bsnm{Seekins}, \binits{J.}},
\oauthor{\bsnm{Mong}, \binits{D.A.}},
\oauthor{\bsnm{Halabi}, \binits{S.S.}},
\oauthor{\bsnm{Sandberg}, \binits{J.K.}},
\oauthor{\bsnm{Jones}, \binits{R.}},
\oauthor{\bsnm{Larson}, \binits{D.B.}},
\oauthor{\bsnm{Langlotz}, \binits{C.P.}},
\oauthor{\bsnm{Patel}, \binits{B.N.}},
\oauthor{\bsnm{Lungren}, \binits{M.P.}},
\oauthor{\bsnm{Ng}, \binits{A.Y.}}:
Chexpert: {A} large chest radiograph dataset with uncertainty labels and expert
  comparison.
CoRR
\textbf{abs/1901.07031}
(2019).
\arxivurl{1901.07031}
\end{botherref}
\endbibitem

%%% 8
\bibitem{Call2021Deep}
\begin{barticle}
\bauthor{\bsnm{Çallı}, \binits{E.}},
\bauthor{\bsnm{Sogancioglu}, \binits{E.}},
\bauthor{\bparticle{van} \bsnm{Ginneken}, \binits{B.}},
\bauthor{\bparticle{van} \bsnm{Leeuwen}, \binits{K.G.}},
\bauthor{\bsnm{Murphy}, \binits{K.}}:
\batitle{Deep learning for chest x-ray analysis: A survey}.
\bjtitle{Medical Image Analysis}
\bvolume{72},
\bfpage{102125}
(\byear{2021}).
doi:\doiurl{10.1016/j.media.2021.102125}
\end{barticle}
\endbibitem

%%% 9
\bibitem{Deng2009ImageNet}
\begin{bchapter}
\bauthor{\bsnm{Deng}, \binits{J.}},
\bauthor{\bsnm{Dong}, \binits{W.}},
\bauthor{\bsnm{Socher}, \binits{R.}},
\bauthor{\bsnm{Li}, \binits{L.-J.}},
\bauthor{\bsnm{Li}, \binits{K.}},
\bauthor{\bsnm{Fei-Fei}, \binits{L.}}:
\bctitle{Imagenet: A large-scale hierarchical image database},
pp. \bfpage{248}--\blpage{255}
(\byear{2009}).
doi:\doiurl{10.1109/CVPR.2009.5206848}.
\bcomment{2009 IEEE Conference on Computer Vision and Pattern Recognition.
  ISSN: 1063-6919}
\end{bchapter}
\endbibitem

%%% 10
\bibitem{Wen2021Rethinking}
\begin{barticle}
\bauthor{\bsnm{Wen}, \binits{Y.}},
\bauthor{\bsnm{Chen}, \binits{L.}},
\bauthor{\bsnm{Deng}, \binits{Y.}},
\bauthor{\bsnm{Zhou}, \binits{C.}}:
\batitle{Rethinking pre-training on medical imaging}.
\bjtitle{Journal of Visual Communication and Image Representation}
\bvolume{78},
\bfpage{103145}
(\byear{2021}).
doi:\doiurl{10.1016/j.jvcir.2021.103145}
\end{barticle}
\endbibitem

%%% 11
\bibitem{Raghu2019Transfusion}
\begin{botherref}
\oauthor{\bsnm{Raghu}, \binits{M.}},
\oauthor{\bsnm{Zhang}, \binits{C.}},
\oauthor{\bsnm{Kleinberg}, \binits{J.}},
\oauthor{\bsnm{Bengio}, \binits{S.}}:
Transfusion: Understanding transfer learning for medical imaging.
arXiv:1902.07208 [cs, stat]
(2019).
arXiv: 1902.07208
\end{botherref}
\endbibitem

%%% 12
\bibitem{Wang2020COVID-Net}
\begin{barticle}
\bauthor{\bsnm{Wang}, \binits{L.}},
\bauthor{\bsnm{Lin}, \binits{Z.Q.}},
\bauthor{\bsnm{Wong}, \binits{A.}}:
\batitle{Covid-net: a tailored deep convolutional neural network design for
  detection of covid-19 cases from chest x-ray images}.
\bjtitle{Scientific Reports}
\bvolume{10}(\bissue{1}),
\bfpage{19549}
(\byear{2020})
\end{barticle}
\endbibitem

%%% 13
\bibitem{Ronneberger2015U-Net}
\begin{bchapter}
\bauthor{\bsnm{Ronneberger}, \binits{O.}},
\bauthor{\bsnm{Fischer}, \binits{P.}},
\bauthor{\bsnm{Brox}, \binits{T.}}:
\bctitle{U-net: Convolutional networks for biomedical image segmentation}.
In: \beditor{\bsnm{Navab}, \binits{N.}},
\beditor{\bsnm{Hornegger}, \binits{J.}},
\beditor{\bsnm{Wells}, \binits{W.M.}},
\beditor{\bsnm{Frangi}, \binits{A.F.}} (eds.)
\bbtitle{Medical Image Computing and Computer-Assisted Intervention -- MICCAI
  2015},
pp. \bfpage{234}--\blpage{241}.
\bpublisher{Springer},
\blocation{Cham}
(\byear{2015})
\end{bchapter}
\endbibitem

%%% 14
\bibitem{Tang2019XLSor}
\begin{bchapter}
\bauthor{\bsnm{Tang}, \binits{Y.-B.}},
\bauthor{\bsnm{Tang}, \binits{Y.-X.}},
\bauthor{\bsnm{Xiao}, \binits{J.}},
\bauthor{\bsnm{Summers}, \binits{R.M.}}:
\bctitle{Xlsor: A robust and accurate lung segmentor on chest x-rays using
  criss-cross attention and customized radiorealistic abnormalities
  generation}.
In: \bbtitle{International Conference on Medical Imaging with Deep Learning},
pp. \bfpage{457}--\blpage{467}
(\byear{2019}).
\bcomment{PMLR}
\end{bchapter}
\endbibitem

%%% 15
\bibitem{Rezatofighi2019Generalized}
\begin{bchapter}
\bauthor{\bsnm{Rezatofighi}, \binits{H.}},
\bauthor{\bsnm{Tsoi}, \binits{N.}},
\bauthor{\bsnm{Gwak}, \binits{J.}},
\bauthor{\bsnm{Sadeghian}, \binits{A.}},
\bauthor{\bsnm{Reid}, \binits{I.}},
\bauthor{\bsnm{Savarese}, \binits{S.}}:
\bctitle{Generalized intersection over union: A metric and a loss for bounding
  box regression}.
In: \bbtitle{Proceedings of the IEEE/CVF Conference on Computer Vision and
  Pattern Recognition},
pp. \bfpage{658}--\blpage{666}
(\byear{2019})
\end{bchapter}
\endbibitem

%%% 16
\bibitem{Bertels2019Optimizing}
\begin{botherref}
\oauthor{\bsnm{Bertels}, \binits{J.}},
\oauthor{\bsnm{Eelbode}, \binits{T.}},
\oauthor{\bsnm{Berman}, \binits{M.}},
\oauthor{\bsnm{Vandermeulen}, \binits{D.}},
\oauthor{\bsnm{Maes}, \binits{F.}},
\oauthor{\bsnm{Bisschops}, \binits{R.}},
\oauthor{\bsnm{Blaschko}, \binits{M.B.}}:
Optimizing the dice score and jaccard index for medical image segmentation:
  Theory and practice.
Medical Image Computing and Computer Assisted Intervention – MICCAI 2019,
92--100
(2019)
\end{botherref}
\endbibitem

%%% 17
\bibitem{Kaiming2019Moco}
\begin{botherref}
\oauthor{\bsnm{He}, \binits{K.}},
\oauthor{\bsnm{Fan}, \binits{H.}},
\oauthor{\bsnm{Wu}, \binits{Y.}},
\oauthor{\bsnm{Xie}, \binits{S.}},
\oauthor{\bsnm{Girshick}, \binits{R.B.}}:
Momentum contrast for unsupervised visual representation learning.
CoRR
\textbf{abs/1911.05722}
(2019).
\arxivurl{1911.05722}
\end{botherref}
\endbibitem

%%% 18
\bibitem{Falcon2020Framework}
\begin{botherref}
\oauthor{\bsnm{Falcon}, \binits{W.}},
\oauthor{\bsnm{Cho}, \binits{K.}}:
A framework for contrastive self-supervised learning and designing a new
  approach.
arXiv:2009.00104 [cs]
(2020).
arXiv: 2009.00104
\end{botherref}
\endbibitem

%%% 19
\bibitem{Sriram2021CovidPrognosis}
\begin{botherref}
\oauthor{\bsnm{Sriram}, \binits{A.}},
\oauthor{\bsnm{Muckley}, \binits{M.J.}},
\oauthor{\bsnm{Sinha}, \binits{K.}},
\oauthor{\bsnm{Shamout}, \binits{F.}},
\oauthor{\bsnm{Pineau}, \binits{J.}},
\oauthor{\bsnm{Geras}, \binits{K.J.}},
\oauthor{\bsnm{Azour}, \binits{L.}},
\oauthor{\bsnm{Aphinyanaphongs}, \binits{Y.}},
\oauthor{\bsnm{Yakubova}, \binits{N.}},
\oauthor{\bsnm{Moore}, \binits{W.}}:
{COVID-19} prognosis via self-supervised representation learning and
  multi-image prediction.
CoRR
\textbf{abs/2101.04909}
(2021).
\arxivurl{2101.04909}
\end{botherref}
\endbibitem

%%% 20
\bibitem{Selvaraju2016gradCAM}
\begin{botherref}
\oauthor{\bsnm{Selvaraju}, \binits{R.R.}},
\oauthor{\bsnm{Das}, \binits{A.}},
\oauthor{\bsnm{Vedantam}, \binits{R.}},
\oauthor{\bsnm{Cogswell}, \binits{M.}},
\oauthor{\bsnm{Parikh}, \binits{D.}},
\oauthor{\bsnm{Batra}, \binits{D.}}:
Grad-cam: Why did you say that? visual explanations from deep networks via
  gradient-based localization.
CoRR
\textbf{abs/1610.02391}
(2016).
\arxivurl{1610.02391}
\end{botherref}
\endbibitem

%%% 21
\bibitem{8438639}
\begin{bchapter}
\bauthor{\bsnm{Xi}, \binits{P.}},
\bauthor{\bsnm{Shu}, \binits{C.}},
\bauthor{\bsnm{Goubran}, \binits{R.}}:
\bctitle{Abnormality detection in mammography using deep convolutional neural
  networks}.
In: \bbtitle{2018 IEEE International Symposium on Medical Measurements and
  Applications (MeMeA)},
pp. \bfpage{1}--\blpage{6}
(\byear{2018}).
doi:\doiurl{10.1109/MeMeA.2018.8438639}
\end{bchapter}
\endbibitem

%%% 22
\bibitem{Stirenko2018Chest}
\begin{botherref}
\oauthor{\bsnm{Stirenko}, \binits{S.}},
\oauthor{\bsnm{Kochura}, \binits{Y.}},
\oauthor{\bsnm{Alienin}, \binits{O.}},
\oauthor{\bsnm{Rokovyi}, \binits{O.}},
\oauthor{\bsnm{Gang}, \binits{P.}},
\oauthor{\bsnm{Zeng}, \binits{W.}},
\oauthor{\bsnm{Gordienko}, \binits{Y.}}:
Chest x-ray analysis of tuberculosis by deep learning with segmentation and
  augmentation.
2018 IEEE 38th International Conference on Electronics and Nanotechnology
  (ELNANO),
422--428
(2018).
doi:\doiurl{10.1109/ELNANO.2018.8477564}.
arXiv: 1803.01199
\end{botherref}
\endbibitem

%%% 23
\bibitem{Candemir2014Lung}
\begin{barticle}
\bauthor{\bsnm{Candemir}, \binits{S.}},
\bauthor{\bsnm{Jaeger}, \binits{S.}},
\bauthor{\bsnm{Palaniappan}, \binits{K.}},
\bauthor{\bsnm{Musco}, \binits{J.P.}},
\bauthor{\bsnm{Singh}, \binits{R.K.}},
\bauthor{\bsnm{Zhiyun~Xue}, \binits{n.}},
\bauthor{\bsnm{Karargyris}, \binits{A.}},
\bauthor{\bsnm{Antani}, \binits{S.}},
\bauthor{\bsnm{Thoma}, \binits{G.}},
\bauthor{\bsnm{McDonald}, \binits{C.J.}}:
\batitle{Lung segmentation in chest radiographs using anatomical atlases with
  nonrigid registration}.
\bjtitle{IEEE transactions on medical imaging}
\bvolume{33}(\bissue{2}),
\bfpage{577}--\blpage{590}
(\byear{2014}).
doi:\doiurl{10.1109/TMI.2013.2290491}.
\bcomment{PMID: 24239990}
\end{barticle}
\endbibitem

%%% 24
\bibitem{Jaeger2014Automatic}
\begin{barticle}
\bauthor{\bsnm{Jaeger}, \binits{S.}},
\bauthor{\bsnm{Karargyris}, \binits{A.}},
\bauthor{\bsnm{Candemir}, \binits{S.}},
\bauthor{\bsnm{Folio}, \binits{L.}},
\bauthor{\bsnm{Siegelman}, \binits{J.}},
\bauthor{\bsnm{Callaghan}, \binits{F.}},
\bauthor{\bsnm{Zhiyun~Xue}, \binits{n.}},
\bauthor{\bsnm{Palaniappan}, \binits{K.}},
\bauthor{\bsnm{Singh}, \binits{R.K.}},
\bauthor{\bsnm{Antani}, \binits{S.}},
\bauthor{\bsnm{Thoma}, \binits{G.}},
\bauthor{\bsnm{Yi-Xiang~Wang}, \binits{n.}},
\bauthor{\bsnm{Pu-Xuan~Lu}, \binits{n.}},
\bauthor{\bsnm{McDonald}, \binits{C.J.}}:
\batitle{Automatic tuberculosis screening using chest radiographs}.
\bjtitle{IEEE transactions on medical imaging}
\bvolume{33}(\bissue{2}),
\bfpage{233}--\blpage{245}
(\byear{2014}).
doi:\doiurl{10.1109/TMI.2013.2284099}.
\bcomment{PMID: 24108713}
\end{barticle}
\endbibitem

%%% 25
\bibitem{AsgariTaghanaki2021Deep}
\begin{barticle}
\bauthor{\bsnm{Asgari~Taghanaki}, \binits{S.}},
\bauthor{\bsnm{Abhishek}, \binits{K.}},
\bauthor{\bsnm{Cohen}, \binits{J.P.}},
\bauthor{\bsnm{Cohen-Adad}, \binits{J.}},
\bauthor{\bsnm{Hamarneh}, \binits{G.}}:
\batitle{Deep semantic segmentation of natural and medical images: a review}.
\bjtitle{Artificial Intelligence Review}
\bvolume{54}(\bissue{1}),
\bfpage{137}--\blpage{178}
(\byear{2021}).
doi:\doiurl{10.1007/s10462-020-09854-1}
\end{barticle}
\endbibitem

%%% 26
\bibitem{Wang2017Segmentation}
\begin{bchapter}
\bauthor{\bsnm{Wang}, \binits{C.}}:
\bctitle{Segmentation of multiple structures in chest radiographs using
  multi-task fully convolutional networks}.
In: \beditor{\bsnm{Sharma}, \binits{P.}},
\beditor{\bsnm{Bianchi}, \binits{F.M.}} (eds.)
\bbtitle{Image Analysis},
pp. \bfpage{282}--\blpage{289}.
\bpublisher{Springer},
\blocation{Cham}
(\byear{2017})
\end{bchapter}
\endbibitem

%%% 27
\bibitem{Cai2020review}
\begin{botherref}
\oauthor{\bsnm{Cai}, \binits{L.}},
\oauthor{\bsnm{Gao}, \binits{J.}},
\oauthor{\bsnm{Zhao}, \binits{D.}}:
A review of the application of deep learning in medical image classification
  and segmentation.
Annals of Translational Medicine
\textbf{8}(11)
(2020).
doi:\doiurl{10.21037/atm.2020.02.44}.
PMID: 32617333 PMCID: PMC7327346
\end{botherref}
\endbibitem

%%% 28
\bibitem{Hesamian2019Deep}
\begin{barticle}
\bauthor{\bsnm{Hesamian}, \binits{M.H.}},
\bauthor{\bsnm{Jia}, \binits{W.}},
\bauthor{\bsnm{He}, \binits{X.}},
\bauthor{\bsnm{Kennedy}, \binits{P.}}:
\batitle{Deep learning techniques for medical image segmentation: Achievements
  and challenges}.
\bjtitle{Journal of Digital Imaging}
\bvolume{32}(\bissue{4}),
\bfpage{582}--\blpage{596}
(\byear{2019}).
doi:\doiurl{10.1007/s10278-019-00227-x}.
\bcomment{PMID: 31144149 PMCID: PMC6646484}
\end{barticle}
\endbibitem

%%% 29
\bibitem{Huang2020CCNet:}
\begin{botherref}
\oauthor{\bsnm{Huang}, \binits{Z.}},
\oauthor{\bsnm{Wang}, \binits{X.}},
\oauthor{\bsnm{Wei}, \binits{Y.}},
\oauthor{\bsnm{Huang}, \binits{L.}},
\oauthor{\bsnm{Shi}, \binits{H.}},
\oauthor{\bsnm{Liu}, \binits{W.}},
\oauthor{\bsnm{Huang}, \binits{T.S.}}:
Ccnet: Criss-cross attention for semantic segmentation.
arXiv:1811.11721 [cs]
(2020).
arXiv: 1811.11721
\end{botherref}
\endbibitem

%%% 30
\bibitem{Shiraishi2000Development}
\begin{barticle}
\bauthor{\bsnm{Shiraishi}, \binits{J.}},
\bauthor{\bsnm{Katsuragawa}, \binits{S.}},
\bauthor{\bsnm{Ikezoe}, \binits{J.}},
\bauthor{\bsnm{Matsumoto}, \binits{T.}},
\bauthor{\bsnm{Kobayashi}, \binits{T.}},
\bauthor{\bsnm{Komatsu}, \binits{K.-i.}},
\bauthor{\bsnm{Matsui}, \binits{M.}},
\bauthor{\bsnm{Fujita}, \binits{H.}},
\bauthor{\bsnm{Kodera}, \binits{Y.}},
\bauthor{\bsnm{Doi}, \binits{K.}}:
\batitle{Development of a digital image database for chest radiographs with and
  without a lung nodule}.
\bjtitle{American Journal of Roentgenology}
\bvolume{174}(\bissue{1}),
\bfpage{71}--\blpage{74}
(\byear{2000}).
doi:\doiurl{10.2214/ajr.174.1.1740071}.
\bcomment{publisher: American Roentgen Ray Society}
\end{barticle}
\endbibitem

%%% 31
\bibitem{Jaeger2014Two}
\begin{barticle}
\bauthor{\bsnm{Jaeger}, \binits{S.}},
\bauthor{\bsnm{Candemir}, \binits{S.}},
\bauthor{\bsnm{Antani}, \binits{S.}},
\bauthor{\bsnm{Wáng}, \binits{Y.-X.J.}},
\bauthor{\bsnm{Lu}, \binits{P.-X.}},
\bauthor{\bsnm{Thoma}, \binits{G.}}:
\batitle{Two public chest x-ray datasets for computer-aided screening of
  pulmonary diseases}.
\bjtitle{Quantitative Imaging in Medicine and Surgery}
\bvolume{4}(\bissue{6}),
\bfpage{475}--\blpage{477}
(\byear{2014}).
doi:\doiurl{10.3978/j.issn.2223-4292.2014.11.20}.
\bcomment{PMID: 25525580 PMCID: PMC4256233}
\end{barticle}
\endbibitem

%%% 32
\bibitem{Huang2018Multimodal}
\begin{botherref}
\oauthor{\bsnm{Huang}, \binits{X.}},
\oauthor{\bsnm{Liu}, \binits{M.-Y.}},
\oauthor{\bsnm{Belongie}, \binits{S.}},
\oauthor{\bsnm{Kautz}, \binits{J.}}:
Multimodal unsupervised image-to-image translation.
arXiv:1804.04732 [cs, stat]
(2018).
arXiv: 1804.04732
\end{botherref}
\endbibitem

%%% 33
\bibitem{Johnson2019MIMIC-CXR}
\begin{botherref}
\oauthor{\bsnm{Johnson}, \binits{A.E.W.}},
\oauthor{\bsnm{Pollard}, \binits{T.J.}},
\oauthor{\bsnm{Berkowitz}, \binits{S.J.}},
\oauthor{\bsnm{Greenbaum}, \binits{N.R.}},
\oauthor{\bsnm{Lungren}, \binits{M.P.}},
\oauthor{\bsnm{Deng}, \binits{C.}},
\oauthor{\bsnm{Mark}, \binits{R.G.}},
\oauthor{\bsnm{Horng}, \binits{S.}}:
{MIMIC-CXR:} {A} large publicly available database of labeled chest
  radiographs.
CoRR
\textbf{abs/1901.07042}
(2019).
\arxivurl{1901.07042}
\end{botherref}
\endbibitem

%%% 34
\bibitem{Santosh2018Automated}
\begin{barticle}
\bauthor{\bsnm{Santosh}, \binits{K.C.}},
\bauthor{\bsnm{Antani}, \binits{S.}}:
\batitle{Automated chest x-ray screening: Can lung region symmetry help detect
  pulmonary abnormalities?}
\bjtitle{IEEE transactions on medical imaging}
\bvolume{37}(\bissue{5}),
\bfpage{1168}--\blpage{1177}
(\byear{2018}).
doi:\doiurl{10.1109/TMI.2017.2775636}.
\bcomment{PMID: 29727280}
\end{barticle}
\endbibitem

%%% 35
\bibitem{Karargyris2016Combination}
\begin{barticle}
\bauthor{\bsnm{Karargyris}, \binits{A.}},
\bauthor{\bsnm{Siegelman}, \binits{J.}},
\bauthor{\bsnm{Tzortzis}, \binits{D.}},
\bauthor{\bsnm{Jaeger}, \binits{S.}},
\bauthor{\bsnm{Candemir}, \binits{S.}},
\bauthor{\bsnm{Xue}, \binits{Z.}},
\bauthor{\bsnm{Santosh}, \binits{K.C.}},
\bauthor{\bsnm{Vajda}, \binits{S.}},
\bauthor{\bsnm{Antani}, \binits{S.}},
\bauthor{\bsnm{Folio}, \binits{L.}},
\bauthor{\bsnm{Thoma}, \binits{G.R.}}:
\batitle{Combination of texture and shape features to detect pulmonary
  abnormalities in digital chest x-rays}.
\bjtitle{International Journal of Computer Assisted Radiology and Surgery}
\bvolume{11}(\bissue{1}),
\bfpage{99}--\blpage{106}
(\byear{2016}).
doi:\doiurl{10.1007/s11548-015-1242-x}.
\bcomment{PMID: 26092662}
\end{barticle}
\endbibitem

%%% 36
\bibitem{Rousseeuw1999Fast}
\begin{barticle}
\bauthor{\bsnm{Rousseeuw}, \binits{P.J.}},
\bauthor{\bsnm{Driessen}, \binits{K.V.}}:
\batitle{A fast algorithm for the minimum covariance determinant estimator}.
\bjtitle{Technometrics}
\bvolume{41}(\bissue{3}),
\bfpage{212}--\blpage{223}
(\byear{1999}).
doi:\doiurl{10.1080/00401706.1999.10485670}.
\bcomment{publisher: Taylor \& Francis}
\end{barticle}
\endbibitem

%%% 37
\bibitem{Cleverley2020role}
\begin{barticle}
\bauthor{\bsnm{Cleverley}, \binits{J.}},
\bauthor{\bsnm{Piper}, \binits{J.}},
\bauthor{\bsnm{Jones}, \binits{M.M.}}:
\batitle{The role of chest radiography in confirming covid-19 pneumonia}.
\bjtitle{BMJ}
\bvolume{370},
\bfpage{2426}
(\byear{2020}).
doi:\doiurl{10.1136/bmj.m2426}.
\bcomment{publisher: British Medical Journal Publishing Group section: Practice
  PMID: 32675083}
\end{barticle}
\endbibitem

%%% 38
\bibitem{Liu2019SDFN}
\begin{barticle}
\bauthor{\bsnm{Liu}, \binits{H.}},
\bauthor{\bsnm{Wang}, \binits{L.}},
\bauthor{\bsnm{Nan}, \binits{Y.}},
\bauthor{\bsnm{Jin}, \binits{F.}},
\bauthor{\bsnm{Wang}, \binits{Q.}},
\bauthor{\bsnm{Pu}, \binits{J.}}:
\batitle{Sdfn: Segmentation-based deep fusion network for thoracic disease
  classification in chest x-ray images}.
\bjtitle{Computerized Medical Imaging and Graphics}
\bvolume{75},
\bfpage{66}--\blpage{73}
(\byear{2019}).
doi:\doiurl{10.1016/j.compmedimag.2019.05.005}.
\bcomment{arXiv: 1810.12959}
\end{barticle}
\endbibitem

%%% 39
\bibitem{chen2020simple}
\begin{botherref}
\oauthor{\bsnm{Chen}, \binits{T.}},
\oauthor{\bsnm{Kornblith}, \binits{S.}},
\oauthor{\bsnm{Norouzi}, \binits{M.}},
\oauthor{\bsnm{Hinton}, \binits{G.}}:
A Simple Framework for Contrastive Learning of Visual Representations
(2020).
\arxivurl{2002.05709}
\end{botherref}
\endbibitem

\end{thebibliography}

\newcommand{\BMCxmlcomment}[1]{}

\BMCxmlcomment{

<refgrp>

<bibl id="B1">
  <title><p>Computer-Aided Diagnosis in Medical Imaging: Historical Review,
  Current Status and Future Potential</p></title>
  <aug>
    <au><snm>Doi</snm><fnm>K</fnm></au>
  </aug>
  <source>Computerized medical imaging and graphics : the official journal of
  the Computerized Medical Imaging Society</source>
  <pubdate>2007</pubdate>
  <volume>31</volume>
  <issue>4-5</issue>
  <fpage>198</fpage>
  <lpage>-211</lpage>
</bibl>

<bibl id="B2">
  <title><p>An automatic computer-aided detection scheme for pneumoconiosis on
  digital chest radiographs</p></title>
  <aug>
    <au><snm>Yu</snm><fnm>P</fnm></au>
    <au><snm>Xu</snm><fnm>H</fnm></au>
    <au><snm>Zhu</snm><fnm>Y</fnm></au>
    <au><snm>Yang</snm><fnm>C</fnm></au>
    <au><snm>Sun</snm><fnm>X</fnm></au>
    <au><snm>Zhao</snm><fnm>J</fnm></au>
  </aug>
  <source>Journal of Digital Imaging</source>
  <pubdate>2011</pubdate>
  <volume>24</volume>
  <issue>3</issue>
  <fpage>382</fpage>
  <lpage>-393</lpage>
  <note>PMID: 20174852 PMCID: PMC3092047</note>
</bibl>

<bibl id="B3">
  <title><p>A review on lung boundary detection in chest X-rays</p></title>
  <aug>
    <au><snm>Candemir</snm><fnm>S</fnm></au>
    <au><snm>Antani</snm><fnm>S</fnm></au>
  </aug>
  <source>International Journal of Computer Assisted Radiology and
  Surgery</source>
  <pubdate>2019</pubdate>
  <volume>14</volume>
  <issue>4</issue>
  <fpage>563</fpage>
  <lpage>-576</lpage>
</bibl>

<bibl id="B4">
  <title><p>Assessment of Convolutional Neural Networks for Automated
  Classification of Chest Radiographs</p></title>
  <aug>
    <au><snm>Dunnmon</snm><fnm>JA</fnm></au>
    <au><snm>Yi</snm><fnm>D</fnm></au>
    <au><snm>Langlotz</snm><fnm>CP</fnm></au>
    <au><snm>Ré</snm><fnm>C</fnm></au>
    <au><snm>Rubin</snm><fnm>DL</fnm></au>
    <au><snm>Lungren</snm><fnm>MP</fnm></au>
  </aug>
  <source>Radiology</source>
  <pubdate>2019</pubdate>
  <volume>290</volume>
  <issue>2</issue>
  <fpage>537</fpage>
  <lpage>-544</lpage>
  <note>publisher: Radiological Society of North America</note>
</bibl>

<bibl id="B5">
  <title><p>Automated abnormality classification of chest radiographs using
  deep convolutional neural networks</p></title>
  <aug>
    <au><snm>Tang</snm><fnm>YX</fnm></au>
    <au><snm>Tang</snm><fnm>YB</fnm></au>
    <au><snm>Peng</snm><fnm>Y</fnm></au>
    <au><snm>Yan</snm><fnm>K</fnm></au>
    <au><snm>Bagheri</snm><fnm>M</fnm></au>
    <au><snm>Redd</snm><fnm>BA</fnm></au>
    <au><snm>Brandon</snm><fnm>CJ</fnm></au>
    <au><snm>Lu</snm><fnm>Z</fnm></au>
    <au><snm>Han</snm><fnm>M</fnm></au>
    <au><snm>Xiao</snm><fnm>J</fnm></au>
    <au><snm>Summers</snm><fnm>RM</fnm></au>
  </aug>
  <source>npj Digital Medicine</source>
  <pubdate>2020</pubdate>
  <volume>3</volume>
  <issue>1</issue>
  <fpage>70</fpage>
</bibl>

<bibl id="B6">
  <title><p>Caveats in Generating Medical Imaging Labels from Radiology
  Reports</p></title>
  <aug>
    <au><snm>Olatunji</snm><fnm>T</fnm></au>
    <au><snm>Yao</snm><fnm>L</fnm></au>
    <au><snm>Covington</snm><fnm>B</fnm></au>
    <au><snm>Rhodes</snm><fnm>A</fnm></au>
    <au><snm>Upton</snm><fnm>A</fnm></au>
  </aug>
  <source>arXiv:1905.02283 [cs, eess]</source>
  <pubdate>2019</pubdate>
  <url>http://arxiv.org/abs/1905.02283</url>
  <note>arXiv: 1905.02283</note>
</bibl>

<bibl id="B7">
  <title><p>CheXpert: {A} Large Chest Radiograph Dataset with Uncertainty
  Labels and Expert Comparison</p></title>
  <aug>
    <au><snm>Irvin</snm><fnm>J</fnm></au>
    <au><snm>Rajpurkar</snm><fnm>P</fnm></au>
    <au><snm>Ko</snm><fnm>M</fnm></au>
    <au><snm>Yu</snm><fnm>Y</fnm></au>
    <au><snm>Ciurea{-}Ilcus</snm><fnm>S</fnm></au>
    <au><snm>Chute</snm><fnm>C</fnm></au>
    <au><snm>Marklund</snm><fnm>H</fnm></au>
    <au><snm>Haghgoo</snm><fnm>B</fnm></au>
    <au><snm>Ball</snm><fnm>RL</fnm></au>
    <au><snm>Shpanskaya</snm><fnm>KS</fnm></au>
    <au><snm>Seekins</snm><fnm>J</fnm></au>
    <au><snm>Mong</snm><fnm>DA</fnm></au>
    <au><snm>Halabi</snm><fnm>SS</fnm></au>
    <au><snm>Sandberg</snm><fnm>JK</fnm></au>
    <au><snm>Jones</snm><fnm>R</fnm></au>
    <au><snm>Larson</snm><fnm>DB</fnm></au>
    <au><snm>Langlotz</snm><fnm>CP</fnm></au>
    <au><snm>Patel</snm><fnm>BN</fnm></au>
    <au><snm>Lungren</snm><fnm>MP</fnm></au>
    <au><snm>Ng</snm><fnm>AY</fnm></au>
  </aug>
  <source>CoRR</source>
  <pubdate>2019</pubdate>
  <volume>abs/1901.07031</volume>
  <url>http://arxiv.org/abs/1901.07031</url>
</bibl>

<bibl id="B8">
  <title><p>Deep learning for chest X-ray analysis: A survey</p></title>
  <aug>
    <au><snm>Çallı</snm><fnm>E</fnm></au>
    <au><snm>Sogancioglu</snm><fnm>E</fnm></au>
    <au><snm>Ginneken</snm><fnm>B</fnm></au>
    <au><snm>Leeuwen</snm><fnm>KG</fnm></au>
    <au><snm>Murphy</snm><fnm>K</fnm></au>
  </aug>
  <source>Medical Image Analysis</source>
  <pubdate>2021</pubdate>
  <volume>72</volume>
  <fpage>102125</fpage>
</bibl>

<bibl id="B9">
  <title><p>ImageNet: A large-scale hierarchical image database</p></title>
  <aug>
    <au><snm>Deng</snm><fnm>J</fnm></au>
    <au><snm>Dong</snm><fnm>W</fnm></au>
    <au><snm>Socher</snm><fnm>R</fnm></au>
    <au><snm>Li</snm><fnm>LJ</fnm></au>
    <au><snm>Li</snm><fnm>K</fnm></au>
    <au><snm>Fei Fei</snm><fnm>L</fnm></au>
  </aug>
  <source>2009 IEEE Conference on Computer Vision and Pattern
  Recognition</source>
  <pubdate>2009</pubdate>
  <fpage>248</fpage>
  <lpage>-255</lpage>
  <note>ISSN: 1063-6919</note>
</bibl>

<bibl id="B10">
  <title><p>Rethinking pre-training on medical imaging</p></title>
  <aug>
    <au><snm>Wen</snm><fnm>Y</fnm></au>
    <au><snm>Chen</snm><fnm>L</fnm></au>
    <au><snm>Deng</snm><fnm>Y</fnm></au>
    <au><snm>Zhou</snm><fnm>C</fnm></au>
  </aug>
  <source>Journal of Visual Communication and Image Representation</source>
  <pubdate>2021</pubdate>
  <volume>78</volume>
  <fpage>103145</fpage>
</bibl>

<bibl id="B11">
  <title><p>Transfusion: Understanding Transfer Learning for Medical
  Imaging</p></title>
  <aug>
    <au><snm>Raghu</snm><fnm>M</fnm></au>
    <au><snm>Zhang</snm><fnm>C</fnm></au>
    <au><snm>Kleinberg</snm><fnm>J</fnm></au>
    <au><snm>Bengio</snm><fnm>S</fnm></au>
  </aug>
  <source>arXiv:1902.07208 [cs, stat]</source>
  <pubdate>2019</pubdate>
  <url>http://arxiv.org/abs/1902.07208</url>
  <note>arXiv: 1902.07208</note>
</bibl>

<bibl id="B12">
  <title><p>COVID-Net: a tailored deep convolutional neural network design for
  detection of COVID-19 cases from chest X-ray images</p></title>
  <aug>
    <au><snm>Wang</snm><fnm>L</fnm></au>
    <au><snm>Lin</snm><fnm>ZQ</fnm></au>
    <au><snm>Wong</snm><fnm>A</fnm></au>
  </aug>
  <source>Scientific Reports</source>
  <pubdate>2020</pubdate>
  <volume>10</volume>
  <issue>1</issue>
  <fpage>19549</fpage>
</bibl>

<bibl id="B13">
  <title><p>U-Net: Convolutional Networks for Biomedical Image
  Segmentation</p></title>
  <aug>
    <au><snm>Ronneberger</snm><fnm>O</fnm></au>
    <au><snm>Fischer</snm><fnm>P</fnm></au>
    <au><snm>Brox</snm><fnm>T</fnm></au>
  </aug>
  <source>Medical Image Computing and Computer-Assisted Intervention -- MICCAI
  2015</source>
  <publisher>Cham: Springer International Publishing</publisher>
  <editor>Navab, Nassir and Hornegger, Joachim and Wells, William M. and
  Frangi, Alejandro F.</editor>
  <pubdate>2015</pubdate>
  <fpage>234</fpage>
  <lpage>-241</lpage>
</bibl>

<bibl id="B14">
  <title><p>Xlsor: A robust and accurate lung segmentor on chest x-rays using
  criss-cross attention and customized radiorealistic abnormalities
  generation</p></title>
  <aug>
    <au><snm>Tang</snm><fnm>YB</fnm></au>
    <au><snm>Tang</snm><fnm>YX</fnm></au>
    <au><snm>Xiao</snm><fnm>J</fnm></au>
    <au><snm>Summers</snm><fnm>RM</fnm></au>
  </aug>
  <source>International Conference on Medical Imaging with Deep
  Learning</source>
  <pubdate>2019</pubdate>
  <fpage>457</fpage>
  <lpage>-467</lpage>
</bibl>

<bibl id="B15">
  <title><p>Generalized intersection over union: A metric and a loss for
  bounding box regression</p></title>
  <aug>
    <au><snm>Rezatofighi</snm><fnm>H</fnm></au>
    <au><snm>Tsoi</snm><fnm>N</fnm></au>
    <au><snm>Gwak</snm><fnm>J</fnm></au>
    <au><snm>Sadeghian</snm><fnm>A</fnm></au>
    <au><snm>Reid</snm><fnm>I</fnm></au>
    <au><snm>Savarese</snm><fnm>S</fnm></au>
  </aug>
  <source>Proceedings of the IEEE/CVF Conference on Computer Vision and Pattern
  Recognition</source>
  <pubdate>2019</pubdate>
  <fpage>658</fpage>
  <lpage>-666</lpage>
</bibl>

<bibl id="B16">
  <title><p>Optimizing the Dice Score and Jaccard Index for Medical Image
  Segmentation: Theory and Practice</p></title>
  <aug>
    <au><snm>Bertels</snm><fnm>J</fnm></au>
    <au><snm>Eelbode</snm><fnm>T</fnm></au>
    <au><snm>Berman</snm><fnm>M</fnm></au>
    <au><snm>Vandermeulen</snm><fnm>D</fnm></au>
    <au><snm>Maes</snm><fnm>F</fnm></au>
    <au><snm>Bisschops</snm><fnm>R</fnm></au>
    <au><snm>Blaschko</snm><fnm>MB</fnm></au>
  </aug>
  <source>Medical Image Computing and Computer Assisted Intervention – MICCAI
  2019</source>
  <publisher>Cham: Springer International Publishing</publisher>
  <editor>Shen, Dinggang and Liu, Tianming and Peters, Terry M. and Staib,
  Lawrence H. and Essert, Caroline and Zhou, Sean and Yap, Pew-Thian and Khan,
  Ali</editor>
  <series><title><p>Lecture Notes in Computer Science</p></title></series>
  <pubdate>2019</pubdate>
  <fpage>92</fpage>
  <lpage>-100</lpage>
</bibl>

<bibl id="B17">
  <title><p>Momentum Contrast for Unsupervised Visual Representation
  Learning</p></title>
  <aug>
    <au><snm>He</snm><fnm>K</fnm></au>
    <au><snm>Fan</snm><fnm>H</fnm></au>
    <au><snm>Wu</snm><fnm>Y</fnm></au>
    <au><snm>Xie</snm><fnm>S</fnm></au>
    <au><snm>Girshick</snm><fnm>RB</fnm></au>
  </aug>
  <source>CoRR</source>
  <pubdate>2019</pubdate>
  <volume>abs/1911.05722</volume>
  <url>http://arxiv.org/abs/1911.05722</url>
</bibl>

<bibl id="B18">
  <title><p>A Framework For Contrastive Self-Supervised Learning And Designing
  A New Approach</p></title>
  <aug>
    <au><snm>Falcon</snm><fnm>W</fnm></au>
    <au><snm>Cho</snm><fnm>K</fnm></au>
  </aug>
  <source>arXiv:2009.00104 [cs]</source>
  <pubdate>2020</pubdate>
  <url>http://arxiv.org/abs/2009.00104</url>
  <note>arXiv: 2009.00104</note>
</bibl>

<bibl id="B19">
  <title><p>{COVID-19} Prognosis via Self-Supervised Representation Learning
  and Multi-Image Prediction</p></title>
  <aug>
    <au><snm>Sriram</snm><fnm>A</fnm></au>
    <au><snm>Muckley</snm><fnm>MJ</fnm></au>
    <au><snm>Sinha</snm><fnm>K</fnm></au>
    <au><snm>Shamout</snm><fnm>F</fnm></au>
    <au><snm>Pineau</snm><fnm>J</fnm></au>
    <au><snm>Geras</snm><fnm>KJ</fnm></au>
    <au><snm>Azour</snm><fnm>L</fnm></au>
    <au><snm>Aphinyanaphongs</snm><fnm>Y</fnm></au>
    <au><snm>Yakubova</snm><fnm>N</fnm></au>
    <au><snm>Moore</snm><fnm>W</fnm></au>
  </aug>
  <source>CoRR</source>
  <pubdate>2021</pubdate>
  <volume>abs/2101.04909</volume>
  <url>https://arxiv.org/abs/2101.04909</url>
</bibl>

<bibl id="B20">
  <title><p>Grad-CAM: Why did you say that? Visual Explanations from Deep
  Networks via Gradient-based Localization</p></title>
  <aug>
    <au><snm>Selvaraju</snm><fnm>RR</fnm></au>
    <au><snm>Das</snm><fnm>A</fnm></au>
    <au><snm>Vedantam</snm><fnm>R</fnm></au>
    <au><snm>Cogswell</snm><fnm>M</fnm></au>
    <au><snm>Parikh</snm><fnm>D</fnm></au>
    <au><snm>Batra</snm><fnm>D</fnm></au>
  </aug>
  <source>CoRR</source>
  <pubdate>2016</pubdate>
  <volume>abs/1610.02391</volume>
  <url>http://arxiv.org/abs/1610.02391</url>
</bibl>

<bibl id="B21">
  <title><p>Abnormality Detection in Mammography using Deep Convolutional
  Neural Networks</p></title>
  <aug>
    <au><snm>Xi</snm><fnm>P</fnm></au>
    <au><snm>Shu</snm><fnm>C</fnm></au>
    <au><snm>Goubran</snm><fnm>R</fnm></au>
  </aug>
  <source>2018 IEEE International Symposium on Medical Measurements and
  Applications (MeMeA)</source>
  <pubdate>2018</pubdate>
  <fpage>1</fpage>
  <lpage>6</lpage>
</bibl>

<bibl id="B22">
  <title><p>Chest X-Ray Analysis of Tuberculosis by Deep Learning with
  Segmentation and Augmentation</p></title>
  <aug>
    <au><snm>Stirenko</snm><fnm>S</fnm></au>
    <au><snm>Kochura</snm><fnm>Y</fnm></au>
    <au><snm>Alienin</snm><fnm>O</fnm></au>
    <au><snm>Rokovyi</snm><fnm>O</fnm></au>
    <au><snm>Gang</snm><fnm>P</fnm></au>
    <au><snm>Zeng</snm><fnm>W</fnm></au>
    <au><snm>Gordienko</snm><fnm>Y</fnm></au>
  </aug>
  <source>2018 IEEE 38th International Conference on Electronics and
  Nanotechnology (ELNANO)</source>
  <pubdate>2018</pubdate>
  <fpage>422</fpage>
  <lpage>-428</lpage>
  <note>arXiv: 1803.01199</note>
</bibl>

<bibl id="B23">
  <title><p>Lung segmentation in chest radiographs using anatomical atlases
  with nonrigid registration</p></title>
  <aug>
    <au><snm>Candemir</snm><fnm>S</fnm></au>
    <au><snm>Jaeger</snm><fnm>S</fnm></au>
    <au><snm>Palaniappan</snm><fnm>K</fnm></au>
    <au><snm>Musco</snm><fnm>JP</fnm></au>
    <au><snm>Singh</snm><fnm>RK</fnm></au>
    <au><snm>Zhiyun Xue</snm><fnm>n</fnm></au>
    <au><snm>Karargyris</snm><fnm>A</fnm></au>
    <au><snm>Antani</snm><fnm>S</fnm></au>
    <au><snm>Thoma</snm><fnm>G</fnm></au>
    <au><snm>McDonald</snm><fnm>CJ</fnm></au>
  </aug>
  <source>IEEE transactions on medical imaging</source>
  <pubdate>2014</pubdate>
  <volume>33</volume>
  <issue>2</issue>
  <fpage>577</fpage>
  <lpage>-590</lpage>
  <note>PMID: 24239990</note>
</bibl>

<bibl id="B24">
  <title><p>Automatic tuberculosis screening using chest
  radiographs</p></title>
  <aug>
    <au><snm>Jaeger</snm><fnm>S</fnm></au>
    <au><snm>Karargyris</snm><fnm>A</fnm></au>
    <au><snm>Candemir</snm><fnm>S</fnm></au>
    <au><snm>Folio</snm><fnm>L</fnm></au>
    <au><snm>Siegelman</snm><fnm>J</fnm></au>
    <au><snm>Callaghan</snm><fnm>F</fnm></au>
    <au><snm>Zhiyun Xue</snm><fnm>n</fnm></au>
    <au><snm>Palaniappan</snm><fnm>K</fnm></au>
    <au><snm>Singh</snm><fnm>RK</fnm></au>
    <au><snm>Antani</snm><fnm>S</fnm></au>
    <au><snm>Thoma</snm><fnm>G</fnm></au>
    <au><snm>Yi Xiang Wang</snm><fnm>n</fnm></au>
    <au><snm>Pu Xuan Lu</snm><fnm>n</fnm></au>
    <au><snm>McDonald</snm><fnm>CJ</fnm></au>
  </aug>
  <source>IEEE transactions on medical imaging</source>
  <pubdate>2014</pubdate>
  <volume>33</volume>
  <issue>2</issue>
  <fpage>233</fpage>
  <lpage>-245</lpage>
  <note>PMID: 24108713</note>
</bibl>

<bibl id="B25">
  <title><p>Deep semantic segmentation of natural and medical images: a
  review</p></title>
  <aug>
    <au><snm>Asgari Taghanaki</snm><fnm>S</fnm></au>
    <au><snm>Abhishek</snm><fnm>K</fnm></au>
    <au><snm>Cohen</snm><fnm>JP</fnm></au>
    <au><snm>Cohen Adad</snm><fnm>J</fnm></au>
    <au><snm>Hamarneh</snm><fnm>G</fnm></au>
  </aug>
  <source>Artificial Intelligence Review</source>
  <pubdate>2021</pubdate>
  <volume>54</volume>
  <issue>1</issue>
  <fpage>137</fpage>
  <lpage>-178</lpage>
</bibl>

<bibl id="B26">
  <title><p>Segmentation of Multiple Structures in Chest Radiographs Using
  Multi-task Fully Convolutional Networks</p></title>
  <aug>
    <au><snm>Wang</snm><fnm>C</fnm></au>
  </aug>
  <source>Image Analysis</source>
  <publisher>Cham: Springer International Publishing</publisher>
  <editor>Sharma, Puneet and Bianchi, Filippo Maria</editor>
  <pubdate>2017</pubdate>
  <fpage>282</fpage>
  <lpage>-289</lpage>
</bibl>

<bibl id="B27">
  <title><p>A review of the application of deep learning in medical image
  classification and segmentation</p></title>
  <aug>
    <au><snm>Cai</snm><fnm>L</fnm></au>
    <au><snm>Gao</snm><fnm>J</fnm></au>
    <au><snm>Zhao</snm><fnm>D</fnm></au>
  </aug>
  <source>Annals of Translational Medicine</source>
  <pubdate>2020</pubdate>
  <volume>8</volume>
  <issue>11</issue>
  <url>https://www.ncbi.nlm.nih.gov/pmc/articles/PMC7327346/</url>
  <note>PMID: 32617333 PMCID: PMC7327346</note>
</bibl>

<bibl id="B28">
  <title><p>Deep Learning Techniques for Medical Image Segmentation:
  Achievements and Challenges</p></title>
  <aug>
    <au><snm>Hesamian</snm><fnm>MH</fnm></au>
    <au><snm>Jia</snm><fnm>W</fnm></au>
    <au><snm>He</snm><fnm>X</fnm></au>
    <au><snm>Kennedy</snm><fnm>P</fnm></au>
  </aug>
  <source>Journal of Digital Imaging</source>
  <pubdate>2019</pubdate>
  <volume>32</volume>
  <issue>4</issue>
  <fpage>582</fpage>
  <lpage>-596</lpage>
  <note>PMID: 31144149 PMCID: PMC6646484</note>
</bibl>

<bibl id="B29">
  <title><p>CCNet: Criss-Cross Attention for Semantic Segmentation</p></title>
  <aug>
    <au><snm>Huang</snm><fnm>Z</fnm></au>
    <au><snm>Wang</snm><fnm>X</fnm></au>
    <au><snm>Wei</snm><fnm>Y</fnm></au>
    <au><snm>Huang</snm><fnm>L</fnm></au>
    <au><snm>Shi</snm><fnm>H</fnm></au>
    <au><snm>Liu</snm><fnm>W</fnm></au>
    <au><snm>Huang</snm><fnm>TS</fnm></au>
  </aug>
  <source>arXiv:1811.11721 [cs]</source>
  <pubdate>2020</pubdate>
  <url>http://arxiv.org/abs/1811.11721</url>
  <note>arXiv: 1811.11721</note>
</bibl>

<bibl id="B30">
  <title><p>Development of a Digital Image Database for Chest Radiographs With
  and Without a Lung Nodule</p></title>
  <aug>
    <au><snm>Shiraishi</snm><fnm>J</fnm></au>
    <au><snm>Katsuragawa</snm><fnm>S</fnm></au>
    <au><snm>Ikezoe</snm><fnm>J</fnm></au>
    <au><snm>Matsumoto</snm><fnm>T</fnm></au>
    <au><snm>Kobayashi</snm><fnm>T</fnm></au>
    <au><snm>Komatsu</snm><fnm>Ki</fnm></au>
    <au><snm>Matsui</snm><fnm>M</fnm></au>
    <au><snm>Fujita</snm><fnm>H</fnm></au>
    <au><snm>Kodera</snm><fnm>Y</fnm></au>
    <au><snm>Doi</snm><fnm>K</fnm></au>
  </aug>
  <source>American Journal of Roentgenology</source>
  <pubdate>2000</pubdate>
  <volume>174</volume>
  <issue>1</issue>
  <fpage>71</fpage>
  <lpage>-74</lpage>
  <note>publisher: American Roentgen Ray Society</note>
</bibl>

<bibl id="B31">
  <title><p>Two public chest X-ray datasets for computer-aided screening of
  pulmonary diseases</p></title>
  <aug>
    <au><snm>Jaeger</snm><fnm>S</fnm></au>
    <au><snm>Candemir</snm><fnm>S</fnm></au>
    <au><snm>Antani</snm><fnm>S</fnm></au>
    <au><snm>Wáng</snm><fnm>YXJ</fnm></au>
    <au><snm>Lu</snm><fnm>PX</fnm></au>
    <au><snm>Thoma</snm><fnm>G</fnm></au>
  </aug>
  <source>Quantitative Imaging in Medicine and Surgery</source>
  <pubdate>2014</pubdate>
  <volume>4</volume>
  <issue>6</issue>
  <fpage>475</fpage>
  <lpage>-477</lpage>
  <note>PMID: 25525580 PMCID: PMC4256233</note>
</bibl>

<bibl id="B32">
  <title><p>Multimodal Unsupervised Image-to-Image Translation</p></title>
  <aug>
    <au><snm>Huang</snm><fnm>X</fnm></au>
    <au><snm>Liu</snm><fnm>MY</fnm></au>
    <au><snm>Belongie</snm><fnm>S</fnm></au>
    <au><snm>Kautz</snm><fnm>J</fnm></au>
  </aug>
  <source>arXiv:1804.04732 [cs, stat]</source>
  <pubdate>2018</pubdate>
  <url>http://arxiv.org/abs/1804.04732</url>
  <note>arXiv: 1804.04732</note>
</bibl>

<bibl id="B33">
  <title><p>{MIMIC-CXR:} {A} large publicly available database of labeled chest
  radiographs</p></title>
  <aug>
    <au><snm>Johnson</snm><fnm>AEW</fnm></au>
    <au><snm>Pollard</snm><fnm>TJ</fnm></au>
    <au><snm>Berkowitz</snm><fnm>SJ</fnm></au>
    <au><snm>Greenbaum</snm><fnm>NR</fnm></au>
    <au><snm>Lungren</snm><fnm>MP</fnm></au>
    <au><snm>Deng</snm><fnm>C</fnm></au>
    <au><snm>Mark</snm><fnm>RG</fnm></au>
    <au><snm>Horng</snm><fnm>S</fnm></au>
  </aug>
  <source>CoRR</source>
  <pubdate>2019</pubdate>
  <volume>abs/1901.07042</volume>
  <url>http://arxiv.org/abs/1901.07042</url>
</bibl>

<bibl id="B34">
  <title><p>Automated Chest X-Ray Screening: Can Lung Region Symmetry Help
  Detect Pulmonary Abnormalities?</p></title>
  <aug>
    <au><snm>Santosh</snm><fnm>K. C.</fnm></au>
    <au><snm>Antani</snm><fnm>S</fnm></au>
  </aug>
  <source>IEEE transactions on medical imaging</source>
  <pubdate>2018</pubdate>
  <volume>37</volume>
  <issue>5</issue>
  <fpage>1168</fpage>
  <lpage>-1177</lpage>
  <note>PMID: 29727280</note>
</bibl>

<bibl id="B35">
  <title><p>Combination of texture and shape features to detect pulmonary
  abnormalities in digital chest X-rays</p></title>
  <aug>
    <au><snm>Karargyris</snm><fnm>A</fnm></au>
    <au><snm>Siegelman</snm><fnm>J</fnm></au>
    <au><snm>Tzortzis</snm><fnm>D</fnm></au>
    <au><snm>Jaeger</snm><fnm>S</fnm></au>
    <au><snm>Candemir</snm><fnm>S</fnm></au>
    <au><snm>Xue</snm><fnm>Z</fnm></au>
    <au><snm>Santosh</snm><fnm>K. C.</fnm></au>
    <au><snm>Vajda</snm><fnm>S</fnm></au>
    <au><snm>Antani</snm><fnm>S</fnm></au>
    <au><snm>Folio</snm><fnm>L</fnm></au>
    <au><snm>Thoma</snm><fnm>GR</fnm></au>
  </aug>
  <source>International Journal of Computer Assisted Radiology and
  Surgery</source>
  <pubdate>2016</pubdate>
  <volume>11</volume>
  <issue>1</issue>
  <fpage>99</fpage>
  <lpage>-106</lpage>
  <note>PMID: 26092662</note>
</bibl>

<bibl id="B36">
  <title><p>A Fast Algorithm for the Minimum Covariance Determinant
  Estimator</p></title>
  <aug>
    <au><snm>Rousseeuw</snm><fnm>PJ</fnm></au>
    <au><snm>Driessen</snm><fnm>KV</fnm></au>
  </aug>
  <source>Technometrics</source>
  <pubdate>1999</pubdate>
  <volume>41</volume>
  <issue>3</issue>
  <fpage>212</fpage>
  <lpage>-223</lpage>
  <note>publisher: Taylor \& Francis</note>
</bibl>

<bibl id="B37">
  <title><p>The role of chest radiography in confirming covid-19
  pneumonia</p></title>
  <aug>
    <au><snm>Cleverley</snm><fnm>J</fnm></au>
    <au><snm>Piper</snm><fnm>J</fnm></au>
    <au><snm>Jones</snm><fnm>MM</fnm></au>
  </aug>
  <source>BMJ</source>
  <pubdate>2020</pubdate>
  <volume>370</volume>
  <fpage>m2426</fpage>
  <note>publisher: British Medical Journal Publishing Group section: Practice
  PMID: 32675083</note>
</bibl>

<bibl id="B38">
  <title><p>SDFN: Segmentation-based Deep Fusion Network for Thoracic Disease
  Classification in Chest X-ray Images</p></title>
  <aug>
    <au><snm>Liu</snm><fnm>H</fnm></au>
    <au><snm>Wang</snm><fnm>L</fnm></au>
    <au><snm>Nan</snm><fnm>Y</fnm></au>
    <au><snm>Jin</snm><fnm>F</fnm></au>
    <au><snm>Wang</snm><fnm>Q</fnm></au>
    <au><snm>Pu</snm><fnm>J</fnm></au>
  </aug>
  <source>Computerized Medical Imaging and Graphics</source>
  <pubdate>2019</pubdate>
  <volume>75</volume>
  <fpage>66</fpage>
  <lpage>-73</lpage>
  <note>arXiv: 1810.12959</note>
</bibl>

<bibl id="B39">
  <title><p>A Simple Framework for Contrastive Learning of Visual
  Representations</p></title>
  <aug>
    <au><snm>Chen</snm><fnm>T</fnm></au>
    <au><snm>Kornblith</snm><fnm>S</fnm></au>
    <au><snm>Norouzi</snm><fnm>M</fnm></au>
    <au><snm>Hinton</snm><fnm>G</fnm></au>
  </aug>
  <pubdate>2020</pubdate>
</bibl>

</refgrp>
} % end of \BMCxmlcomment
% for author-year bibliography (bmc-mathphys or spbasic)
% a) write to bib file (bmc-mathphys only)
% @settings{label, options="nameyear"}
% or include bibliography directly:
% \begin{thebibliography}
% \bibitem{b1}
% \end{thebibliography}
%%%%%%%%%%%%%%%%%%%%%%%%%%%%%%%%%%%
%%                               %%
%% Figures                       %%
%%                               %%
%% NB: this is for captions and  %%
%% Titles. All graphics must be  %%
%% submitted separately and NOT  %%
%% included in the Tex document  %%
%%                               %%
%%%%%%%%%%%%%%%%%%%%%%%%%%%%%%%%%%%
%\end{multicols}
\end{document}